\documentclass[twocolumn,showpacs,preprintnumbers,amsmath,amssymb]{revtex4}
\usepackage{epsfig}
\usepackage{bbm}

\newcommand {\be}{\begin{equation}}
\newcommand {\ee}{\end{equation}}
\newcommand {\bea}{\begin{eqnarray}}
\newcommand {\ea}{\end{eqnarray*}}
\newcommand {\ba}{\begin{eqnarray*}}
\newcommand {\eea}{\end{eqnarray}}

\newcommand {\ham} {{\mathcal H}}

\newcommand{\bm}[1]{ \mbox{\boldmath $#1$}  }

\begin{document}
\title{Breakup of three particles within the 
adiabatic expansion method.}

\author{E. Garrido}
\affiliation{Instituto de Estructura de la Materia, CSIC, Serrano 123, E-28006 Madrid, Spain}
\author{A. Kievsky and M. Viviani}
\affiliation{Istituto Nazionale di Fisica Nucleare, Largo Pontecorvo 3, 56100 Pisa, Italy}

\begin{abstract}
General expressions for the breakup cross sections in the lab frame 
for $1+2$ reactions are given in terms of the hyperspherical adiabatic basis. 
The three-body wave function is expanded in this basis
and the corresponding hyperradial functions are obtained by solving a set
of second order differential equations. The ${\cal S}$-matrix is computed by
using two recently derived integral relations. Even though the method is shown 
to be well suited to describe $1+2$ processes, there are nevertheless
particular configurations in the breakup channel (for example those in which 
two particles move away close to each other in a relative zero-energy state) that need a huge 
number of basis states. This pathology manifests itself in the extremely slow
convergence of the breakup amplitude in terms of the hyperspherical harmonic basis used 
to construct the adiabatic channels. To overcome this difficulty the breakup amplitude 
is extracted from an integral relation as well. For the sake of illustration, 
we consider neutron-deuteron scattering. The results are compared to the available 
benchmark calculations.
\end{abstract}

\pacs{03.65.Nk, 25.10.+s, 31.15.xj}

\maketitle

\section{Introduction}

The use of the hyperspherical adiabatic (HA) expansion method \cite{nie01} to describe a 
$1+2$ collision between a particle and a bound state of two particles, 
called dimer in atomic physics or the deuteron in nuclear physics,
was considered to be rather inefficient due to the slow pattern of convergence of the elastic
channel \cite{bar09}. This fact seemed to limit the applicability of the method only to the 
description of bound states.
The problem is due to the fact that the hyperradial coordinate used in the adiabatic expansion 
is not suitable to describe the asymptotic behavior of the elastic outgoing wave. In this 
case the use of the relative coordinate between the two particles in the outgoing dimer, 
and the one between the center of mass of the dimer and the third
particle is much more convenient. In particular, due to the finite size of the dimer,
this last relative coordinate and the hyperradius coincide only at infinity. For this reason, 
a correct description of the asymptotic 1+2 wave function and, therefore the determination
of the ${\cal S}$-matrix, requires knowledge of the wave function at very large distances, 
which in turn requires a huge amount of terms in the adiabatic expansion \cite{bar09b}.

A new method, based on two integral relations derived from the Kohn variational 
principle (KVP), was introduced in \cite{bar09b} in order to permit the determination of the 
elastic ${\cal S}$-matrix from the internal part of the wave 
function. Therefore, when used together with the adiabatic expansion method, the number of 
adiabatic terms needed in the calculation is much lower. In fact, the pattern of convergence 
is similar to that observed for bound states \cite{bar09b}. The details of the procedure 
for elastic, inelastic and recombination processes are given in Refs.\cite{rom11,kie10} for energies 
below the dimer breakup threshold. The extension to treat the elastic channel at energies 
above the breakup threshold was discussed in Ref.\cite{gar12}.
 
The knowledge of the elastic, inelastic and recombination ${\cal S}$-matrix elements
can be used to compute different observables characterizing the reaction. 
Furthermore, using the unitary condition
\begin{equation}
\sum_m^M |S_{im}|^2+ \sum_n^\infty |{\cal S}_{in}|^2 =1 \,\, ,
\end{equation}
where $m$ indicates the finite set $M$ of the elastic, inelastic and recombination channels and 
$n$ labels the infinite set of breakup channels,
it is possible to obtain the total breakup cross section
\begin{equation}
\sigma_b=\frac{\pi}{k^2}(1-\sum_m^M|S_{im}|^2)
\label{crossb}
\end{equation}
where, for three particles with equal mass, 
we have that $k^2=(4/3)E_i(\hbar^2/m)$ and $E_i$ is the incident 
energy in the center of mass frame. 
Examples of this procedure can be found in Refs.\cite{gar12,gar13} where the inelasticities
in $n-d$ scattering have been computed as well as the recombination and dissociation rates
in the direct and inverse atomic processes 
$ ^4{\rm He}+ {}^4{\rm He}+ {}^4{\rm He}\rightarrow {}^4{\rm He}_2+ {}^4{\rm He}$.
Therefore, an accurate calculation of the elastic and, if allowed, inelastic and recombination 
${\cal S}$-matrix elements leads to an accurate determination of the corresponding total
breakup cross section through the unitary condition. 

On the other hand, when the knowledge of the breakup amplitude is required, the 
$S_{in}$ matrix elements have to be computed explicitly. Using  the HA
method, the index $n$ is related to the number of adiabatic channels
taken into account in the description of the three-body scattering wave function.
In the present work each adiabatic channel is expanded in the hyperspherical harmonic
basis (HH). Therefore a study of the convergence properties of the breakup amplitude in
terms of the adiabatic channels and the HH basis is in order. 
These two convergencies have to be achieved separately.  As we will see, 
the number of adiabatic channels needed to describe the elastic channel with high accuracy
is sufficient for an accurate determination of the breakup amplitude. However
there are particular kinematic conditions of the outgoing particles in which
the convergence of the breakup amplitude is terms of the HH basis becomes very 
delicate. This is the case when two of the particles move away close to each other with
almost zero relative energy. In order to treat this specific configuration we will make
use of an integral relation for the breakup amplitude as discussed for example
in Ref.~\cite{payne00}.

The method discussed in the present work is general and can be applied to
different kinds of three-body reactions. In this work we shall consider the
$N-d$ case, which is frequently described using the Faddeev 
equations, as shown for instance in Refs.\cite{wgreport,screport} or
in the recent review of Ref.\cite{car14},
or using the HH formalism in conjunction
with the KVP~\cite{kiev01}. Calculation using the Faddeev equations
in momentum space including the Coulomb force between the two protons
can be found in Ref.~\cite{deltuva05}. This and the HH method have been
compared in the elastic channel up to $65$ MeV~\cite{dfk05}. Attempts to
explicitly determine the breakup amplitude using the HH formalism
can be found in Refs.~\cite{kiev97,kiev99} whereas the formalism using the
KVP is discussed in Ref.~\cite{viv01}. Those calculations have shown the
intrinsic difficulties of a variational description of the breakup amplitude.
As mentioned above, the main problem appears in the description of particular 
kinematics as for example the case in which two particles, for instance
two neutrons travel close to each in a relative zero-energy state. This configuration 
represents a kind of clusterization inside the breakup amplitude and requires a huge
number of basis states to be described properly. 

Using the experience gained in Refs~\cite{rom11,kie10,gar12} in which the HA method was used
to describe elastic, inelastic and recombination processes by means of two
integral relations derived from the KVP, in the present work we extend
the HA method to describe the breakup amplitude explicitly.
Expressions for the differential cross sections in the
lab frame for $1+2$ reactions at incident energies above the dimer breakup threshold
are derived.
The neutron-deuteron ($n-d$) reaction is studied with a semi-realistic $s$-wave force
for illustration. This choice is motivated by the existence of benchmark calculations 
\cite{fri95} which allow a test of the method. This study is a first step in the
application of the method to describe $N-d$ scattering using realistic two- and three-body
forces and including the Coulomb interaction.

In the first part of the paper we provide the details of the formalism used to compute the 
differential cross sections. This part is divided into several subsections where we
give the expression of the cross section in the lab frame, the expansion 
of the transition amplitude in terms of HA functions and, finally, 
an integral relation to compute the transition amplitude which removes
the convergence problem inherent to the HA expansion. In
Section III the computation of the integral relation is discussed,
in particular the treatment of the long tail of the kernel.
The results obtained for the case of neutron-deuteron breakup are described in Section IV
and the last section is devoted to some conclusive remarks. For the sake of
completeness, the paper includes six appendices in which some derivations not
essential for the understanding of the paper are given. In particular, several 
technical aspects of the scattering 
theory are discussed in terms of the hyperspherical adiabatic expansion method.

\section{Formalism}

In this section we summarize the formalism employed to compute the differential cross sections
for 1+2 reactions at energies above the breakup threshold. To this end, we have divided the section 
into four parts, which correspond to:

$A)$ description of the notation used;

$B)$ derivation of the general expression of the cross section in the lab frame in terms of 
the transition amplitude;

$C)$ expansion of the transition amplitude in terms of the HA functions;

$D)$ derivation of the integral relation in terms of the three-nucleon
scattering wave function in which the outgoing six-dimensional wave is not expanded in 
the HA basis. 

This relation can correct some inaccuracies in the computed
transition amplitude (see for example Ref.\cite{payne00}). It also manifests
the variational character of the method. The details of 
how to compute this matrix element and, in particular, how to compute the long tail 
of the integral contained in this matrix element are discussed in the next section.

Some theoretical derivations, not crucial for an understanding of the formalism,
but in order to have a compact presentation of the method, have 
been collected in the appendices.

\subsection{Notation and coordinates}

Let us denote by $\bm{r}_i$ ($i=1,2,3$) the coordinates of the three particles involved in 
the 1+2 reaction under investigation, and by $\bm{p}_i$ ($i=1,2,3$) their corresponding 
momenta. From these coordinates we construct the usual Jacobi coordinates, which are given by:
\begin{eqnarray}
\bm{x}_i&=&\sqrt{\frac{\mu_{x_i}}{m}} (\bm{r}_j-\bm{r}_k)=\sqrt{\frac{\mu_{x_i}}{m}} \bm{r}_{x_i} \label{eq1}\\ 
\bm{y}_i&=&\sqrt{\frac{\mu_{y_i}}{m}} \left(\bm{r}_i-\frac{m_j \bm{r}_j+m_k \bm{r}_k}{m_j+m_k }\right)=
\sqrt{\frac{\mu_{y_i}}{m}} \bm{r}_{y_i} \label{eq2}, 
\end{eqnarray}
where $\mu_{x_i}$ is the reduced mass of the $jk$ two-body system, $\mu_{y_i}$ is the reduced 
mass of particle $i$ and the two-body system $jk$, $m$ is an arbitrary normalization mass, 
and $m_i$ ($i=1,2,3$) are the masses of the three particles. Cyclic permutations of 
$\{i,j,k\}$ give the three possible sets of Jacobi coordinates.

The corresponding Jacobi coordinates in momentum space take the form:
\begin{eqnarray}
\bm{k}_{x_i}&=&\sqrt{\frac{m}{\mu_{x_i}}} 
   \left( \frac{m_k}{m_j+m_k}\bm{p}_j - \frac{m_j}{m_j+m_k}\bm{p}_k \right) \nonumber\\ 
           &=& \sqrt{\frac{m}{\mu_{x_i}}} \bm{p}_{x_i} \label{eq3} \\ 
\bm{k}_{y_i}&=&\sqrt{\frac{m}{\mu_{y_i}}} 
   \left(\frac{(m_j+m_k)\bm{p}_i}{m_i+m_j+m_k}- \frac{m_i (\bm{p}_j+\bm{p}_k ) }{m_i+m_j+m_k } \right)
   \nonumber \\
   &=&\sqrt{\frac{m}{\mu_{y_i}}} \bm{p}_{y_i}.
\label{eq4}
\end{eqnarray}

From the Jacobi coordinates we construct the hyperspherical coordinates. They are given by 
one radial coordinate, the hyperradius $\rho$, defined as $\rho=\sqrt{x^2+y^2}$ (the 
definition is independent of the Jacobi set used), and five hyperangles, which are given by 
$\alpha_i=\arctan{x_i/y_i}$, and the polar and azimuthal angles describing the direction of 
$\bm{x}_i$ and $\bm{y}_i$, i.e.,
$\Omega_{x_i}\equiv \{\theta_{x_i},\varphi_{x_i}\}$, and 
$\Omega_{y_i}\equiv \{\theta_{y_i},\varphi_{y_i}\}$.
The hyperangles depends on the Jacobi set chosen to describe the three-body system, and we 
shall denote them in a compact form as 
$\Omega_i \equiv \{\alpha_i,\Omega_{x_i},\Omega_{y_i}\}$.

The corresponding hyperspherical coordinates in momentum space are given by 
the three-body momentum $\kappa=\sqrt{k_x^2+k_y^2}$, and the five hyperangles 
$\Omega_{\kappa_i}\equiv \{\alpha_{\kappa_i},\Omega_{k_{x_i}},\Omega_{k_{y_i}} \}$,
where $\alpha_{\kappa_i}=\arctan{k_{x_i}/k_{y_i}}$.
The three-body momentum $\kappa$ is related to the total three-body energy $E$ of the process by 
the expression $\kappa=\sqrt{2mE}/\hbar$.  

Note that the volume element is given in terms of the relative coordinates 
$\bm{r}_{x_i}$ and $\bm{r}_{y_i}$ defined in 
Eqs.(\ref{eq1}) and (\ref{eq2}). Therefore:
\begin{eqnarray}
dV_i=d\bm{r}_{x_i} d\bm{r}_{y_i} &=& \left(\frac{m}{\mu_{x_i}} \right)^{3/2}
\left(\frac{m}{\mu_{y_i}} \right)^{3/2} d\bm{x}_i d\bm{y}_i \label{eq5} \\
&=& \left(\frac{m}{\mu_{x_i}} \right)^{3/2} \left(\frac{m}{\mu_{y_i}} \right)^{3/2} 
\rho^5 d\rho d\Omega_i,
\nonumber
\end{eqnarray}
where $d\Omega_i=\sin^2 \alpha_i \cos^2 \alpha_i d\alpha_i d\Omega_{x_i} d\Omega_{y_i}$, and
which means that the hypersurface element of the hypersphere with hyperradius $\rho$ is given
by: 
\begin{equation}
d\Sigma_i=\left(\frac{m}{\mu_{x_i}} \right)^{3/2} \left(\frac{m}{\mu_{y_i}} \right)^{3/2} 
\rho^5 d\Omega_i .
\label{eq6}
\end{equation}

It is important to note that asymptotically the hyperangles in coordinate ($\Omega_i$) and 
momentum ($\Omega_{\kappa_i}$) space coincide. This is related to the fact that the 
hyperspherical harmonics transform into themselves after a Fourier transformation. A more 
intuitive way of checking this fact is that asymptotically, at a given time $t$, the 
coordinate of particle $i$ is just given by $\bm{r}_i \rightarrow t\bm{p}_i/m_i$.
When replacing this expressions for $\bm{r}_i$, $\bm{r}_j$, and $\bm{r}_k$ into 
Eqs.(\ref{eq1}) and (\ref{eq2}), and taking into account the definitions (\ref{eq3}) and 
(\ref{eq4}), we immediately get that $\bm{x}_i \rightarrow t \bm{k}_{x_i}/m$ and 
$\bm{y}_i \rightarrow t \bm{k}_{y_i}/m$. Therefore,
asymptotically, the polar and azimuthal angles describing the directions of $\bm{x}_i$ and
$\bm{y}_i$ are the same as those describing the directions of $\bm{k}_{x_i}$ and
$\bm{k}_{y_i}$, and also, $x_i/y_i=k_{x_i}/k_{y_i}$ which implies that, asymptotically, 
$\alpha_i=\alpha_{\kappa_i}$. 
Therefore, asymptotically, $d\Omega_i=d\Omega_{\kappa_i}$.

When describing the incoming 1+2 channel it is convenient to choose the Jacobi set such that 
the relative coordinate $\bm{r}_{x}$ in Eq.(\ref{eq1}) is the relative coordinate between the two 
particles in the dimer. In this case the Jacobi momentum $\bm{k}_y$ in Eq.(\ref{eq4}) is given by
$\bm{k}_y=\sqrt{m/\mu_y}\bm{p}_y$ where  
$\bm{p}_y$ is just the incident relative projectile-dimer momentum in the center of mass frame. 
We shall denote these two vectors, $\bm{k}_y$ and $\bm{p}_y$, as $\bm{k}_y^{(in)}$ and $\bm{p}_y^{(in)}$, 
such that they can be distinguished from the corresponding momenta in the final state. The
momentum $\bm{p}_y^{(in)}$ is related to the incident energy $E_{in}$ (in the center of mass frame) 
by the expression $p_y^{(in)}=\sqrt{2\mu_y E_{in}}/\hbar$, and the total energy $E$ is then given 
by $E=E_{in}+E_d$, where 
$E_d$ is the binding energy of the dimer ($E_d<0$).
In the following, unless explicitly mentioned, we shall use this Jacobi set (the
dimer wave function depends only on the $\bm{x}$ coordinate) and we shall omit the index $i$
when referring to the $(\bm{x},\bm{y})$ or $(\bm{k}_x,\bm{k}_y)$ coordinates defined
in Eqs.(\ref{eq1}) to (\ref{eq4}).

\subsection{Breakup cross section in the lab frame}

The differential cross section $d\sigma$ after a 1+2 breakup reaction is given by the 
outgoing flux of the particles through an element of the hypersurface, Eq.(\ref{eq6}), 
normalized with the incident flux.

The expression for the outgoing flux is derived in Appendix \ref{app1}, and it is given by 
Eq.(\ref{apen7}):
\begin{equation}
\mbox{outgoing flux}= \hbar \frac{\kappa}{m} |A_{\sigma_d \sigma_p}^{\sigma_i \sigma_j \sigma_k}|^2
\left(\frac{m}{\mu_x} \right)^{3/2} \left(\frac{m}{\mu_y} \right)^{3/2} d\Omega_\kappa,
\label{eq7}
\end{equation}
where $A_{\sigma_d \sigma_p}^{\sigma_i \sigma_j \sigma_k}$ is the breakup transition amplitude, 
in which we have made explicit the spin projections $\sigma_i$, $\sigma_j$, $\sigma_k$, 
$\sigma_d$, and $\sigma_p$ which correspond to the three particles found after the breakup 
(with spins $s_i$, $s_j$, and $s_k$), and to the dimer (with spin $s_d$) and the projectile 
(with spin $s_p$).  

The incoming flux is the one corresponding to a particle-dimer two-body process, 
and it is given by:
\begin{equation}
\mbox{incoming flux}= \hbar \frac{p_y^{(in)}}{\mu_y}=\hbar \sqrt{\frac{m}{\mu_y}} \frac{k_y^{(in)}}{m},
\label{eq8}
\end{equation}
where the connection between incident momentum $p_y^{(in)}$ and $k_y^{(in)}$ is given in
Eq.(\ref{eq4}).

The ratio between Eqs.(\ref{eq7}) and (\ref{eq8}) gives then the differential cross section in the three-body
center of mass frame, and it takes the form:
\begin{eqnarray}
\frac{d^5\sigma}{d\Omega_\kappa}&=& 
\frac{\kappa}{k_y^{(in)}} \sqrt{\frac{\mu_y}{m}}
\left(\frac{m}{\mu_x}\right)^{3/2} \left(\frac{m}{\mu_y}\right)^{3/2}
 \label{eq9}  \\ & & \times
\frac{1}{(2s_d+1)(2s_p+1)} \sum_{\sigma_i \sigma_j \sigma_k} \sum_{\sigma_d \sigma_p}
\left|
A^{\sigma_i \sigma_j \sigma_k}_{\sigma_d \sigma_p}
\right|^2, 
\nonumber
\end{eqnarray}
where we have already averaged over the initial states and summed over all the possible final 
states. This procedure gives the $1/(2s_d+1)(2s_p+1)$ factor and the summation over all the 
spin projections.

In appendix \ref{app2} we have derived the phase space in terms of the center of mass 
coordinates, Eq.(\ref{apb5}), and in terms of the laboratory coordinates, Eq.(\ref{apb24}). 
For simplicity, derivation of Eq.(\ref{apb24}) has been made assuming three particles with 
equal mass $m$ (which is also taken to be the normalization mass in Eqs.(\ref{eq1}) to 
(\ref{eq4})). Thus the expression below is valid only for this particular case, 
although the generalization to three particles with different masses is
straightforward. Making equal Eqs.(\ref{apb5}) and (\ref{apb24}) we then obtain:
\begin{equation}
\frac{d\Omega_\kappa}{dS d\hat{\bm{p}}_i d\hat{\bm{p}}_j}= \frac{m}{\hbar^2}
\left(\frac{m}{\mu_x}\right)^{3/2} \left(\frac{m}{\mu_y}\right)^{3/2}
\frac{K_S}{\kappa^4},
\label{eq10}
\end{equation}
where $i$ and $j$ refer to two of the outgoing particles, 
$d\hat{\bm{p}}_i=\sin\theta_{p_i}d\theta_{p_i}d\varphi_{p_i}$, 
$\{\theta_{p_i},\varphi_{p_i}\}$
are the polar and azimuthal angles giving the direction of momentum $\bm{p}_i$ 
(and similarly for particle $j$), the arclength $S$ is defined by Eq.(\ref{apb21b}), 
and, finally, $K_S$ is given by Eq.(\ref{apb25}). Since:
\begin{equation}
\frac{d^5\sigma}{dS d\hat{\bm{p}}_i d\hat{\bm{p}}_j}=
\frac{d\Omega_\kappa}{dS d\hat{\bm{p}}_i d\hat{\bm{p}}_j}
\frac{d^5\sigma}{d\Omega_\kappa},
\end{equation}
we get the following final expression for the cross section in the laboratory frame:
\begin{equation}
\frac{d^5\sigma}{dS d\hat{\bm p}_id\hat{\bm p}_j}=
\left(\frac{m}{\mu_x}\right)^{3/2}
\left(\frac{m}{\mu_y}\right)^{3/2}
\frac{m}{\hbar^2} \frac{K_S}{\kappa^4} \frac{d^5\sigma}{d\Omega_\kappa},
\label{eq12}
\end{equation} 
where $d^5\sigma/d\Omega_\kappa$ is given in Eq.(\ref{eq9}). 

In this work we shall focus on neutron-deuteron breakup reactions. For this case we have 
three spin 1/2 particles with mass $m$ equal to the nucleon mass, and such that the spin 
of the dimer and the projectile are, respectively, $s_d=1$ and $s_p=1/2$. Also, 
$k_y^{(in)}$ is given by $\sqrt{3/2} p_y^{(in)}$, where $p_y^{(in)}$ is the relative momentum between 
projectile and dimer. Using then Eqs.(\ref{eq9}) and (\ref{eq12}) we obtain for the $n-d$ 
case:
\begin{equation}
\frac{d^5\sigma}{dS d\hat{\bm p}_id\hat{\bm p}_j}=
\frac{3m}{\hbar^2}
\frac{K_S}{\kappa^3 p_y^{(in)}} 
\sum_{\sigma_i \sigma_j \sigma_j} \sum_{\sigma_d \sigma_p}
\left|
A^{\sigma_i \sigma_j \sigma_k}_{\sigma_d \sigma_p}
\right|^2.
\label{eq13}
\end{equation}

In our calculations the particles $i$ and $j$ in the equation above will be taken to be the two 
outgoing neutrons. The input will be the neutron incident energy in the lab frame 
($E_{in}^{(lab)}$), the polar angles $\theta_{p_i}$ and $\theta_{p_j}$ for the two outgoing 
neutrons, and $\Delta \varphi=\varphi_{p_i}-\varphi_{p_j}$. These angles, as shown in 
Eqs.(\ref{apb11}), (\ref{apb12}), and (\ref{apb13}), determine the values of $\mu_i$, 
$\mu_j$, and $\mu$ entering in $K_S$ (Eq.(\ref{apb25})).

The input incident energy $E_{in}^{(lab)}$ immediately provides the momentum of the projectile 
in the lab frame: 
\begin{equation}
p_y^{(lab)}=\sqrt{2mE_{in}^{(lab)} }/\hbar, 
\label{eq14}
\end{equation}
which also enters in $K_S$.

The lab energy $E_{in}^{(lab)}$ can be easily related to the incident energy in the center of 
mass frame ($E_{in}$), which for the case of the $n-d$ reaction becomes 
$E_{in}=2E_{in}^{(lab)}/3$. From it we can get the center of mass relative momentum $p_y^{(in)}$ 
entering in Eq.(\ref{eq13}), which is given by $p_y^{(in)}=\sqrt{2 \mu_y E_{in}}/\hbar$, where 
$\mu_y=2m/3$ is the projectile-dimer reduced mass.

The cross section given in Eq.(\ref{eq13}) is a function of the arclength $S$. For each 
value of $S$, given the input $E_{in}^{(lab)}$, $\theta_{p_i}$, $\theta_{p_j}$, and 
$\Delta \varphi$, the values of $p_i$ and $p_j$ are uniquely determined, as shown in 
appendix \ref{app3}. These two momenta, together with the already known values of  
$p_y^{(lab)}$, $\mu_i$, $\mu_j$ and $\mu$, permit now to compute $K_S$ according to 
Eq.(\ref{apb25}) and then obtain the differential cross section. 
The only piece remaining is the determination of the breakup transition amplitude 
$A^{\sigma_i \sigma_j \sigma_k}_{\sigma_d \sigma_p}$, which is discussed
in the following sections.

\subsection{The breakup transition amplitude in the HA expansion method}
\label{amplitude}

When using hyperspherical coordinates the three-body hamiltonian operator 
$\hat{\ham}$ takes the form \cite{nie01}:
\begin{equation}
\hat{\ham} =  -\frac{\hbar^2}{2 m} \hat{T}_\rho + \hat {\cal H}_\Omega    ,
\end{equation} 
where $\hat{T}_\rho=\partial^2/\partial\rho^2+(5/\rho)\partial/\partial\rho$ 
is the hyperradial kinetic energy operator and $\hat {\cal H}_\Omega$ is defined as
\begin{equation}
\label{omegaH}
\hat {\cal H}_\Omega = \frac{\hbar^2}{2 m} \frac{\hat{L}^2(\Omega)}{\rho^2}
+\sum_{i<j} V(i,j)
\end{equation}
where $\hat{L}^2(\Omega)$ is the grand-angular operator, and $m$ is an arbitrary normalization mass.
The operator $\hat {\cal H}_\Omega$
contains all the dependence on the hyperangles and the potential
energy, which has been taken to include two-body forces only. Eventually three-body forces can be 
considered as well. This hamiltonian can be solved for fixed values of $\rho$, such that the 
angular eigenfunctions $\Phi_n(\rho,\Omega)$ satisfy
\begin{equation}
\hat { \cal H}_\Omega \Phi^{JM}_n(\rho,\Omega)=\frac{\hbar^2}{2 m} \frac{1}{\rho^2}
\lambda_n(\rho) \Phi^{JM}_n(\rho,\Omega).
\label{eq16}
\end{equation}

The set of angular eigenfunctions $\{\Phi^{JM}_n(\rho,\Omega)\}$ forms the HA basis
with definite values of the total angular momentum and projection $JM$. They form a
complete basis that can be used to expand the three-body wave function. The advantage 
of this basis is that the large distance behavior of each term can be related to the 
different open channels. In particular, the possible elastic, inelastic and recombination
1+2 channels are associated to specific adiabatic terms~\cite{gar12},
whose corresponding eigenvalues $\lambda_n(\rho)$ go asymptotically as $2mE_d\rho^2/\hbar^2$,
where $E_d$ is now the binding energy of the dimer in that specific 1+2 channel.
The remaining infinitely many adiabatic basis terms describe three free particles 
in the continuum, and each of their corresponding eigenvalues $\lambda_n(\rho)$ behaves at 
large distances as $K(K+4)$ where $K$ is the grand-angular quantum number. 
Therefore, 
each breakup channel is associated to a single value of $K$. In other words,
if a breakup angular eigenfunction $\Phi^{JM}_n(\rho,\Omega)$ is expanded in terms of 
the HH basis, at very large distances, only the HH basis elements with that specific value
of $K$ survive. Asymptotically the HA basis describing the breakup channels
coincides with the HH basis since, in this case, the HA basis elements are eigenfunctions 
of the $\hat {\cal H}_\Omega$ operator given in Eq.(\ref{omegaH}) without the interaction term.

In the following we shall focus on a $1+2$ process where inelastic or
recombination channels are not possible.
This means that there is only one possible dimer in the $1+2$ system that can be either the
target or the projectile not having any bound excited state. 
The inclusion of additional $1+2$ channels does not present any intrinsic 
difficulties \cite{rom11}, and they are not relevant for the description 
of the breakup channel discussed in this work. In fact, the $n-d$ reaction that we shall consider
corresponds precisely to this kind of reactions (the deuteron does not have excited states and is 
the only two-nucleon bound system).  

In the appendix \ref{app4} we have shown that the adiabatic expansion of the outgoing three-body 
wave function describing the breakup of the dimer can be written as (Eq.(\ref{expan})):
\begin{eqnarray}
\Psi_{\sigma_d \sigma_p}^{\sigma_i \sigma_j \sigma_k} &=&
(2\pi)^{3/2} \sum_{JM} \sum_{n>1} \frac{1}{(\kappa \rho)^{5/2}} f^J_{n1}(\rho)
  \label{eq17} \\ & & \times
\langle\sigma_i \sigma_j \sigma_j|\Phi_{n}^{JM}(\rho, \Omega_\rho)\rangle 
\langle \sigma_d \sigma_p | \Phi^{JM}_1(\kappa,\Omega_\kappa)\rangle^*,
\nonumber
\end{eqnarray}
where $\Phi^{JM}_1(\kappa,\Omega_\kappa)$ is the adiabatic angular function (in momentum space)
associated to the incident 1+2 channel (channel 1), and $J$ and $M$ are the total angular momentum 
and projection of the three-body (projectile+target) system.
The summation over $n$ refers to all the breakup adiabatic
channels, whose corresponding angular functions are given by
$\Phi_{n}^{JM}(\rho, \Omega_\rho)$ (as mentioned above, we have assumed that the incident
channel 1 is the only 1+2 channel in the three-body system).
The wave function is projected over the initial
spin states $|\sigma_d \sigma_p\rangle$ of target and projectile, and the final spin
states $| \sigma_i \sigma_j \sigma_k\rangle$ of the three particles after the breakup.

The hyperradial functions $f^J_{n1}(\rho)$
in Eq.(\ref{eq17}) depend on the total angular momentum $J$ and
are obtained by solving the set of coupled differential equations
\begin{eqnarray}
\lefteqn{ \hspace*{-1cm}
\left[ -\frac{d^2}{d\rho^2} +  \frac{\lambda_n(\rho)+\frac{15}{4}}{\rho^2} -\frac{2mE}{\hbar^2}
 \right] f^J_{n1}(\rho)} \nonumber \\ & &
- \sum_{n} \left(2 P_{n n'}(\rho) \frac{d}{d\rho} + Q_{n n'}(\rho) \right)f^J_{n'1}(\rho)
= 0,  
\label{radeq}
\end{eqnarray}
where the eigenvalues $\lambda_n(\rho)$ in  Eq.(\ref{eq16}) enter as effective potentials, and
where the functions of the hyperradius $P_{nn'}$ and $Q_{nn'}$ couple the different adiabatic
terms. Details are given in \cite{rom11,gar12}.

For the breakup channels ($n>1$) the radial wave functions $f^J_{n1}(\rho)$ behave asymptotically 
as \cite{cob98}:
\begin{equation}
f^J_{n1}(\rho) \rightarrow \frac{\sqrt{\kappa \rho}}{2} {\cal S}_{1n}^{J} H_{K+2}^{(1)}(\kappa \rho)
\rightarrow \frac{1}{2}i^{-K} {\cal S}_{1n}^{J} \sqrt{\frac{2}{\pi}} e^{i \frac{3\pi}{4}} e^{i \kappa \rho},
\end{equation}
where $K$ is the grand-angular quantum number associated to the breakup adiabatic channel $n$, 
$H_{K+2}^{(1)}$ is a Hankel function of first kind, and ${\cal S}_{1n}^J$ is the 
corresponding matrix element of the ${\cal S}$-matrix.

Using the expression above, we can write the asymptotic behavior of the outgoing wave 
function in Eq.(\ref{eq17}) as:
\begin{equation}
\Psi_{\sigma_d \sigma_n}^{\sigma_1 \sigma_2 \sigma_3}\rightarrow 
\frac{e^{i \kappa \rho}}{\rho^{5/2}} A_{\sigma_d \sigma_n}^{\sigma_1 \sigma_2 \sigma_3},
\end{equation}
where the breakup transition amplitude is given by:
\begin{eqnarray}
\lefteqn{
A_{\sigma_d \sigma_p}^{\sigma_i \sigma_j \sigma_k}  = 
\frac{2\pi}{\kappa^{5/2}} e^{i \frac{3\pi}{4}}
}
\label{eq20} \\ & &
\times \sum_{JM}  \sum_{n>1} i^{-K} {\cal S}_{1n}^J
\langle \sigma_i \sigma_j \sigma_k | \Phi_n^{JM}(\Omega_\rho)\rangle
\langle \sigma_d \sigma_p | \Phi^{JM}_1(\kappa,\Omega_\kappa)\rangle^*,
\nonumber
\end{eqnarray}
and $\Phi_n^{JM}(\Omega_\rho)=\lim_{\rho \rightarrow \infty} \Phi_n^{JM}(\rho,\Omega_\rho)$.

The asymptotic behavior of the angular eigenfunction $\Phi^{JM}_1(\kappa,\Omega_\kappa)$ is derived
in appendix~\ref{app5}, and for the particular case of relative $s$-waves between the particles takes 
the form (Eq.(\ref{apd10})):
\begin{eqnarray}
\lefteqn{ \hspace*{-1cm}
\langle \sigma_d \sigma_p | \Phi^{JM}_1(\kappa,\Omega_\kappa)\rangle^* \rightarrow } \nonumber \\ & &
\frac{1}{\sqrt{4\pi}} \left( \frac{\mu_x}{m} \right)^{3/4} \frac{\kappa^2}{\sqrt{k_y^{(in)}}}
\langle s_d \sigma_d s_p \sigma_p | J M \rangle,
\label{eq22}
\end{eqnarray}
where $k_y^{(in)}$ is related to the incident relative projectile-dimer momentum $p_y^{(in)}$ through
Eq.(\ref{eq4}) and $\mu_x$ is the reduced mass of the two particles in the dimer.
The expression above permits to write the transition amplitude given in Eq.(\ref{eq20}) as:
\begin{eqnarray}
\lefteqn{
A_{\sigma_d \sigma_n}^{\sigma_1 \sigma_2 \sigma_3}=
\frac{\sqrt{\pi}}{\sqrt{\kappa k_y^{(in)}}}
\left( \frac{\mu_x}{m} \right)^{3/4} e^{i \frac{3\pi}{4}}
}\label{eq23} \\ & & 
\sum_{JM} \langle s_d \sigma_d s_p \sigma_p |JM \rangle \sum_{n>1} i^{-K} {\cal S}_{1n}^J
\langle \sigma_1 \sigma_2 \sigma_3 | \Phi_n^{JM}(\Omega_\rho)\rangle,
\nonumber
\end{eqnarray}
which is valid for relative $s$-waves only. 

The angular eigenfunctions $\Phi_n^{JM}(\rho,\Omega_\rho)$
can be decomposed in the three Faddeev amplitudes
as (see Refs.\cite{nie01,bar09} for details)
\begin{eqnarray}
\lefteqn{ \hspace*{-3mm} \Phi_n^{JM}(\rho,\Omega)=}  \\ & &
    \Phi_n^{(i)JM}(\rho,\Omega_i)+ 
              \Phi_n^{(j)JM}(\rho,\Omega_j)+
                        \Phi_n^{(k)JM}(\rho,\Omega_k),
\nonumber
\end{eqnarray}
where each of the three components of the angular eigenfunction is written in terms of each 
of the three possible sets of Jacobi coordinates. Moreover,
in the present work the angular eigenfunctions $\Phi_n^{JM}(\rho,\Omega_i)$ are 
expanded in terms of the HH basis. In general, the corresponding expansion coefficients 
contain the dependence of the angular eigenfunction on $\rho$. Asymptotically each 
breakup adiabatic channel is associated to some specific value of $K$, 
and the eigenfunction becomes a $\rho$-independent expansion. More precisely, it results in
a linear combination of HH functions having well-defined grand-angular quantum number. 
In particular, if we consider only relative $s$-waves between the 
particles the following expression can be obtained
\begin{eqnarray}
\lefteqn{
\langle \sigma_i \sigma_j \sigma_k | \Phi_n^{(i)JM}(\Omega_i)\rangle=}
\label{eq21} \\ & &
\frac{1}{4\pi}\sum_{s_{x_{i}}}
C_{K s_{x_{i}}}^{(n)} N_K P_\nu^{(\frac{1}{2} \frac{1}{2})}(\cos 2\alpha_i)
\langle \sigma_i \sigma_j \sigma_k | \chi_{s_{x_{i}} s_{y_{i}}}^{JM} \rangle,
\nonumber
\end{eqnarray}
where the $C$'s are the coefficients in the expansion, $N_K$ is a normalization coefficient and
$P_\nu^{(\frac{1}{2}\frac{1}{2})}$ is a Jacobi polynomial. Moreover $K$ is associated to the
asymptotic behavior of the adiabatic channel,
$\nu=K/2$, $s_x$ is the coupling between the spins of the two particles used to construct the 
$\bm{x}$ Jacobi coordinate, $s_y$ is the spin of the third particle, and $\chi_{s_x s_y}^{JM}$ 
is the total three-body spin function arising from the coupling of $s_x$ and
$s_y$ to the total angular momentum $J$ with projection $M$.
Taking this into account, 
the transition amplitude in Eq.(\ref{eq23}) can be written in a more compact form as:
\begin{eqnarray}
\lefteqn{
A_{\sigma_d \sigma_p}^{\sigma_i \sigma_j \sigma_k}=} \label{eq25} \\ & &
\sum_{JM} \langle s_d \sigma_d s_p \sigma_p |JM \rangle
\sum_{q=1}^3 \sum_{s_{x_q}} A_{s_{x_q}}(q) 
\langle \sigma_i \sigma_j \sigma_k | \chi_{s_{x_q} s_{y_q}}^{JM} \rangle,
\nonumber
\end{eqnarray}
where the index $q$ numbers the three possible sets of Jacobi coordinates and
\begin{eqnarray}
A_{s_{x_q}}(q)&=&\frac{1}{4\sqrt{\pi}} \frac{1}{\sqrt{\kappa k_y^{(in)}}}
\left( \frac{\mu_x}{m} \right)^{3/4} e^{i \frac{3\pi}{4}} 
\label{eq26} \\ & & \times
\sum_{n>1} i^{-K} C_{K s_{x_q}}^{(n)} {\cal S}_{1n}^J N_K P_\nu^{(\frac{1}{2} \frac{1}{2})}(\cos 2\alpha_q),
\nonumber
\end{eqnarray} 
where we have made use of Eq.(\ref{eq21}).
From Eq.(\ref{eq25}) we can then finally write:
\begin{eqnarray}
\sum_{\sigma_p \sigma_d} \sum_{\sigma_i \sigma_j \sigma_k} 
\left|  A_{\sigma_d \sigma_p}^{\sigma_i \sigma_j \sigma_k}  \right|^2&=&
\sum_J (2J+1) 
\label{eq27} \\ & & \hspace*{-3.5cm}
\times \left\{
\sum_{p,q=1}^{3} \sum_{s_{x_p} s_{x_q}}
A_{s_{x_p}}^*(p) A_{s_{x_q}}(q)
\langle \chi^J_{s_{x_p} s_{y_p}}|\chi^J_{s_{x_q} s_{y_q}} \rangle
\right\},
\nonumber
\end{eqnarray}
where $p$ and $q$ run over the three possible sets of Jacobi coordinates, and 
$| \chi^J_{s_{x_p} s_{y_p}}\rangle$ is the three-body spin function in the Jacobi set $p$, 
where the spin $s_{x_p}$, associated to the Jacobi coordinate ${\bm x}_p$, couples to the 
spin $s_{y_p}$ of the third particle to give the total three-body angular momentum $J$ (all 
the orbital angular momenta are assumed to be zero). Finally, Eq.(\ref{eq27}), together with
Eq.(\ref{eq26}), permits to obtain the cross section in the center of mass frame in
Eq.(\ref{eq9}), and therefore the cross section in the lab frame as given by Eq.(\ref{eq12}).

It should be noticed that the transition amplitude in Eq.(\ref{eq23}) is obtained from the 
asymptotic behavior of the wave function in Eq.(\ref{eq17}). This means that the hyperangles 
$\Omega_\rho$ entering in Eq.(\ref{eq23}), or (\ref{eq26}), are the asymptotic hyperangles, 
which are known to be the same in coordinate and momentum space. 

\subsection{Integral relation for the breakup transition amplitude}
\label{2d}

The working equation in the calculation of the breakup amplitude using the HA
basis is Eq.(\ref{eq26}). The feasibility of the method could be limited by the
number of adiabatic terms needed in the expansion, given by the index $n$ which in
principle runs up to $\infty$. Since asymptotically the
HA and the HH basis tend to be the same, the coefficients $C_{K s_{x_q}}^{(n)}$ have 
the property of being close to a non-zero constant when 
$K$ takes the value associated to the asymptotic behavior of the adiabatic 
potential $n$, and close to zero otherwise. Therefore the number of terms in the expansion is 
intrinsically
related to the ability of the HH basis to describe the asymptotic configurations. 

If the hyperangle $\alpha_q=\arctan(k_{x_q}/k_{y_q})$ approaches zero, 
the relative momentum of the two particles connected by the $\bm{x}_q$ Jacobi coordinate 
approaches zero as well, and the two particles appear in an almost
zero-energy relative state. As mentioned before, this produces a kind of clusterization
in the breakup amplitude very difficult to describe with the HH basis. In fact,
for this particular geometry, the breakup adiabatic angular eigenfunctions entering in the 
transition amplitude (see Eq.(\ref{eq23})) should be similar to the one corresponding to a 1+2 channel,
given by Eq.(\ref{apd1}), but replacing 
the bound dimer wave function by the zero energy two-body wave function.  

\begin{figure}
\centerline{\psfig{figure=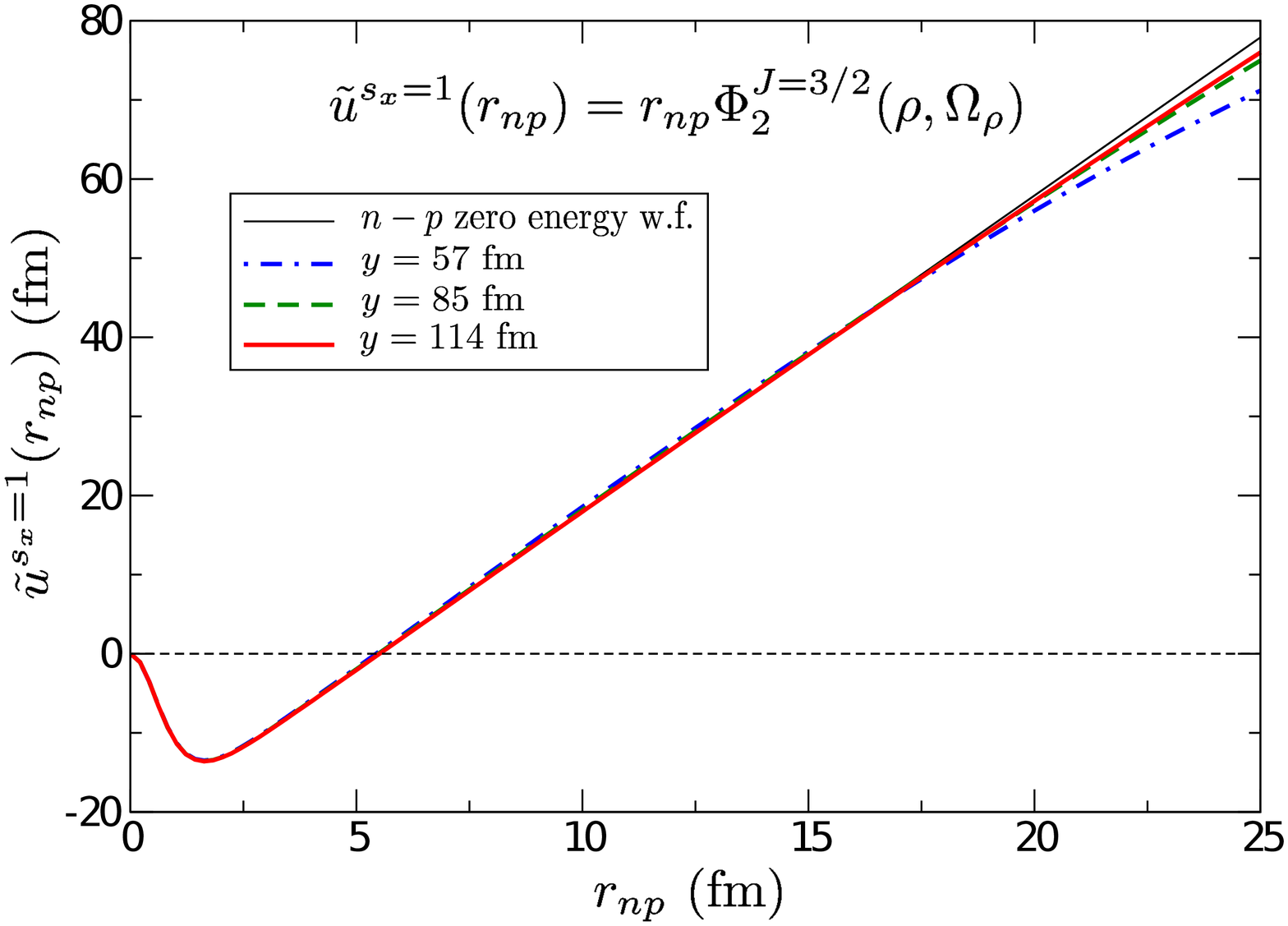,width=8.5cm,angle=0}}
\caption{(Color online) Wave function $\tilde{u}^{s_x=1}(r_{np})$ obtained from the lowest
breakup angular eigenfunction $\Phi_2^{J=1/2}(\rho,\Omega_\rho)$ for the $n-d$ case
with total angular momentum $J=3/2$ (see text), and for three different fixed values
of the Jacobi coordinate $y$, 57 fm (thick dot-dashed), 85 fm (thick dashed), and 114 fm (thick solid).
The thin solid line is the zero energy neutron-proton wave function $u^{s_x=1}$ introduced
in Eq.(\ref{0en}).  }
\label{fig1}
\end{figure}

To illustrate this point we take the $n-d$ case with total spin $J=3/2$ as
an example. In this case, according to Eq.(\ref{apd1}), for the geometries having
$\alpha \approx 0$ ($x \ll y$) and assuming relative $s$-waves 
between the particles we should get:
\begin{equation}
\Phi_n^{J=3/2}(\rho,\Omega_\rho)\stackrel{\rho\rightarrow \infty}{\rightarrow}
        \frac{u^{s_x=1}}{r_{np}}
\label{0en}
\end{equation}
for all the angular eigenfunctions associated to breakup channels ($n>1$), where 
$u^{s_x=1}/r_{np}$ is the zero energy neutron-proton wave function, and $r_{np}$ is the relative
distance between the neutron and the proton. In 
Fig.\ref{fig1} we plot the function 
$\tilde{u}^{s_x=1}(r_{np})=r_{np} \Phi_2^{J=3/2}(\rho, \Omega_\rho)$
as a function $r_{np}$ for three fixed values of the Jacobi coordinate $y$, 57 fm (dot-dashed curve), 
85 fm (dashed curve), and 114 fm (solid curve). Obviously, the larger the value of $y$ the smaller
the value of the hyperangle $\alpha$ associated to a given $r_{np}$, and therefore the more should
the function $\tilde{u}^{s_x=1}$ approach the zero-energy two-body function $u^{s_x=1}$ given 
in Eq.(\ref{0en}), which is shown in Fig.\ref{fig1} by the thin solid line. 
The expected behavior is what we observe in the figure, where, as we can see, for $y=114$ fm the 
function $\tilde{u}^{s_x=1}$ (thick solid curve) matches the two-body wave function $u^{s_x=1}$ 
(thin solid curve) pretty well up to almost 20 fm (the thin solid curve has been scaled to fit the same 
minimum as $\tilde{u}^{s_x=1}$). 

The zero-energy two-body wave function is linear in the relative radial coordinate
(see Fig.\ref{fig1}), and therefore proportional to $\sin \alpha_q$. 
The reconstruction of this behavior by use of an expansion in terms of $\cos 2\alpha_q$, as in 
the expansion given in Eq.(\ref{eq26}), requires in principle infinitely many terms.
Accordingly, as $\alpha_q\rightarrow 0$ the number of adiabatic terms needed to get a convergent
value for $A_{s_{x_p}}(p)$ increases without limit. Hence,  
Eq.(\ref{eq26}) is not operative in this particular situation.

To overcome this problem we shall develop in this section an alternative expression for the 
transition amplitude where the adiabatic expansion in Eq.(\ref{eq23}), in terms of the 
coefficients $C_{K s_{x_q}}^{(n)}$, does not enter explicitly. 
This procedure will be more expensive from a numerical point of view, 
but more accurate in the kinematic regions where 
$\alpha_\kappa \approx 0$. The starting point here is the well known expansion of the 
three-body plane wave in terms of the hyperspherical harmonics:
\begin{eqnarray}
\lefteqn{
  e^{i({\small \bm{k}}_{x}\cdot{\small \bm{x}}+{\small \bm{k}}_{y}\cdot{\small \bm{y}})} 
|\sigma_i \sigma_j \sigma_k \rangle
= 
} \label{eq30} \\ & &
\frac{(2\pi)^3}{(\kappa \rho)^2} \sum_{JM} \sum_{[K]} i^K J_{K+2}(\kappa \rho) 
{\cal Y}_{[K]}^{JM}(\Omega_\rho)
 \langle \sigma_i \sigma_j \sigma_k | {\cal Y}_{[K]}^{JM*}(\Omega_\kappa),
\nonumber
\end{eqnarray}
where ${\cal Y}_{[K]}^{JM}$ is a HH function coupled to a three-body spin function \cite{nie01}. 
All the quantum numbers are collected into the set $[K] \equiv\{K,\ell_x,\ell_y,L,s_x,S\}$. 
On the other hand, the angular eigenfunctions in the 
adiabatic expansion corresponding to breakup channels ($n>1$) are, asymptotically,
linear combination of hyperspherical harmonics. The two bases can be formally related as
\begin{equation}
{\cal Y}_{[K]}^{JM}(\Omega)=\sum_{n>1} \langle \Phi^{JM}_n(\Omega) | {\cal Y}_{[K]}^{JM}(\Omega) \rangle 
\Phi^{JM}_n(\Omega),
\end{equation}
with the sum over $n$ restricted to those channels associated to the grand-angular
quantum number $K$. Replacing in Eq.(\ref{eq30}) we obtain
\begin{eqnarray}
\lefteqn{
  e^{i({\small \bm{k}}_{x}\cdot{\small \bm{x}}+{\small \bm{k}}_{y}\cdot{\small \bm{y}})} 
                  |\sigma_i \sigma_j \sigma_k\rangle  
= 
}  \label{eq31} \\ & &
\frac{(2\pi)^3}{(\kappa \rho)^2} \sum_{JM} \sum_{n>1} i^K J_{K+2}(\kappa \rho) 
\Phi_{n}^{JM}(\Omega_\rho) \langle \sigma_i \sigma_j \sigma_k | \Phi_n^{JM}(\Omega_\kappa)\rangle^*,
\nonumber
\end{eqnarray}
where we have used that 
$\sum_{[K]} \langle \Phi^{JM}_n | {\cal Y}_{[K]}^{JM} \rangle
 \langle {\cal Y}_{[K]}^{JM} | \Phi^{JM}_{n'} \rangle =\delta_{nn'}$.

As shown in Ref.\cite{gar12}, the regular outgoing wave functions for the breakup channels are given by:
\begin{equation}
F_n=\sqrt{\frac{\pi}{2}}\frac{1}{\rho^2} J_{K+2}(\kappa \rho) \Phi_n^{JM}(\Omega_\rho).
\label{eq36a}
\end{equation}

Therefore Eq.(\ref{eq31}) can be written in terms of $F_n$, and in particular it can be used 
to write the following matrix element: 
\begin{eqnarray}
\lefteqn{
\langle \Psi_1^{JM}|\hat{\cal H}-E | 
  e^{i({\small \bm{k}}_{x}\cdot{\small \bm{x}}+{\small \bm{k}}_{y}\cdot{\small \bm{y}})} 
                 |\sigma_i \sigma_j \sigma_k\rangle   \rangle= }
 \label{eq32} \\ & &
\frac{(2\pi)^3}{\kappa^2} \sqrt{\frac{2}{\pi}} \sum_{n>1} i^K
\langle \Psi_1^{JM} | \hat{\cal H}-E | F_n \rangle 
\langle \sigma_i  \sigma_j  \sigma_k | \Phi_n^{JM}(\Omega_\kappa)\rangle^*,
\nonumber
\end{eqnarray}
where $\hat{\cal H}$ is the three-body hamiltonian, $E$ is the total three-body energy, and 
$\Psi_1^{JM}$ is the three-body wave function corresponding to the incoming 1+2 channel labeled
by 1.

In Ref.\cite{gar12} it was also shown that the ${\cal K}$-matrix for a breakup process can be 
obtained through
two integral relations that provide the two matrices $A$ and $B$, such that 
the ${\cal K}$-matrix of the reaction takes the form ${\cal K}=-{A}^{-1}B$. 
In particular, the $ij$-term of each of these two matrices is given by:
\begin{eqnarray}
A_{ij}&=& -\frac{2m}{\hbar^2} \langle \Psi_i^{JM} |\hat{\cal H}-E | G_j\rangle \\
B_{ij}&=&  \frac{2m}{\hbar^2} \langle \Psi_i^{JM} |\hat{\cal H}-E | F_j\rangle \label{eqb},
\end{eqnarray}
where $G_j$ is defined as $F_j$, Eq.(\ref{eq36a}), but replacing the regular Bessel function
$J_{K+2}$ by the irregular Bessel function $Y_{K+2}$.

Use of Eq.(\ref{eqb}) permits to write Eq.(\ref{eq32}) as:
\begin{eqnarray}
\lefteqn{ \hspace*{-5mm}
\langle \Psi_1^{JM}|\hat{\cal H}-E | 
  e^{i({\small \bm{k}}_{x}\cdot{\small \bm{x}}+{\small \bm{k}}_{y}\cdot{\small \bm{y}})} 
|\sigma_i \sigma_j \sigma_k\rangle   \rangle= 
} \label{eq34}\\  & &
\frac{(2\pi)^3}{\kappa^2} \sqrt{\frac{2}{\pi}} \frac{\hbar^2}{2m} \sum_{n>1} i^{K} B_{1n}
\langle \sigma_i  \sigma_j  \sigma_k | \Phi_n^{JM}(\Omega_\kappa)\rangle^*,
\nonumber
\end{eqnarray}
where it is important to keep in mind that $1$ refers to the incoming channel (1+2 channel) and
$n$ refers to the breakup channels (therefore $n>1$).

Note that the matrix element in Eq.(\ref{eq34}) is just the first component of a column vector 
whose $i$-th term is given by $\langle \Psi_i^{JM}|\hat{\cal H}-E | \mbox{plane wave}\rangle$
($i=1,\cdots,N$, where $N$ is the number of adiabatic channels included in
the calculation) where for $i>2$ the wave functions $\Psi_i^{JM}$ describe scattering processes with
three ingoing particles. If we multiply from the left such column vector by any $N\times N$ matrix $M$, 
the result would be a new column vector whose first term would be given by Eq.(\ref{eq34}) but 
replacing $B_{1n}$ by $\left(M\cdot B\right)_{1n}$. Therefore, if we take $M=A^{-1}$, the first component 
of the new vector will be given by Eq.(\ref{eq34}) but replacing $B_{1n}$ by $(A^{-1}B)_{1n}$, which 
is nothing but $-{\cal K}_{1n}$.

For the same reason, since ${\cal K}=-i({\cal S}+\mathbbm{I})^{-1}({\cal S}-\mathbbm{I})$, 
if we multiply from the left the new column vector by $({\cal S}+\mathbbm{I})$, the new matrix
element in Eq.(\ref{eq34}) would be given not in terms of ${\cal K}_{1n}$, but in terms of 
$({\cal S}-\mathbbm{I})_{1n}$, which reduces to ${\cal S}_{1n}$ in our case, where $1$ 
refers to the $1+2$ incident channel and $n$ corresponds to outgoing breakup channels ($n>1$).

Summarizing, the first step is to compute the $A$ and 
$B$ matrices as described in Ref.\cite{gar12}, from which the 
${\cal S}-$matrix of the reaction can be obtained. Successively a proper normalized
scattering state is constructed according to:
\begin{equation}
\Psi^{JM}\rightarrow ({\cal S}+\mathbbm{I})A^{-1} \Psi^{JM},
\end{equation}
and once this is done the matrix element given in Eq.(\ref{eq34}) transforms into:
\begin{eqnarray}
\lefteqn{ \hspace*{-5mm}
\langle \Psi_1^{JM}|\hat{\cal H}-E | 
  e^{i({\small \bm{k}}_{x}\cdot{\small \bm{x}}+{\small \bm{k}}_{y}\cdot{\small \bm{y}})} 
|\sigma_i \sigma_j \sigma_k\rangle   \rangle=
} \label{eq36} \\ & &
 i \frac{(2\pi)^3}{\kappa^2} \sqrt{\frac{2}{\pi}} \frac{\hbar^2}{2m} \sum_{n>1} i^{-K}
{\cal S}_{1n} \langle \sigma_i  \sigma_j  \sigma_k | \Phi_n^{JM}(\Omega_\kappa)\rangle^*,
\nonumber
\end{eqnarray}
where the replacement of $i^K$ by $i^{-K}$ is irrelevant, since the difference is a factor 
$i^{2K}=(-1)^K$, which is either 1 for positive parity states ($K$ even for all $n$) or $-1$ 
for negative parity states ($K$ odd for all $n$). In any case, this is not playing any role, 
since the calculation of the cross section will contain the square of the matrix element above.

In Eq.(\ref{eq36}) the angular function 
$\langle \sigma_i  \sigma_j  \sigma_k | \Phi_n(\Omega_\kappa)\rangle$ is just 
the Fourier transform of Eq.(\ref{eq21}), whose analytical form is exactly the 
same as in Eq.(\ref{eq21}) but with the angles understood in momentum space 
(asymptotically the hyperangles in coordinate and momentum space are the same).

Comparing Eqs.(\ref{eq23}) and (\ref{eq36}), we can identify the breakup transition amplitude as:
\begin{eqnarray}
\lefteqn{
A_{\sigma_d \sigma_n}^{\sigma_i \sigma_j \sigma_k}=  
\frac{\pi}{\sqrt{2}}
\frac{\kappa^2}{\sqrt{\kappa k_y}} \frac{1}{(2\pi)^3}
\left( \frac{\mu_x}{m} \right)^{3/4} e^{i \frac{\pi}{4}} \frac{2m}{\hbar^2}
}\label{eq37} \\ & &
\times \sum_{JM} \langle s_d \sigma_d s_p \sigma_p |JM \rangle 
\langle \Psi_1^{JM}|\hat{\cal H}-E | 
  e^{i({\small \bm{k}}_{x}\cdot{\small \bm{x}}+{\small \bm{k}}_{y}\cdot{\small \bm{y}})} 
             |\sigma_i \sigma_j \sigma_k\rangle   \rangle. 
\nonumber
\end{eqnarray}

Since the plane wave is the solution of the free hamiltonian, the matrix 
element in Eq.(\ref{eq36}) is actually the sum of three matrix elements, each of them involving one of 
the three two-body potentials. In turn, each of these three matrix elements is
more easily treated in the Jacobi set such that the potential depends on the 
$\bm{x}$ coordinate only.
In other words, we can write:
\begin{eqnarray}
\lefteqn{\hspace*{-1cm}
\langle \Psi_1^{JM}|\hat{\cal H}-E | 
  e^{i({\small \bm{k}}_{x}\cdot{\small \bm{x}}+{\small \bm{k}}_{y}\cdot{\small \bm{y}})} 
                       |\sigma_i \sigma_j \sigma_k\rangle   \rangle=  
} \nonumber \\ & &
\sum_{q=1}^3 
\langle \Psi_1^{JM}|V_q(x_q) | 
  e^{i({\small \bm{k}}_{x_q}\cdot{\small \bm{x}_q}+{\small \bm{k}}_{y_q}\cdot{\small \bm{y}_q})} 
                       |\sigma_i \sigma_j \sigma_k\rangle   \rangle,  
\label{eq38}
\end{eqnarray}
where $V_q(x_q)$ is the two-body potential between the two particles connected by the Jacobi coordinate
$\bm{x}_q$.

If we now consider that
\begin{eqnarray}
\lefteqn{
\langle \Psi_1^{JM}|V_q(x_q) | 
  e^{i({\small \bm{k}}_{x_q}\cdot{\small \bm{x}_q}+{\small \bm{k}}_{y_q}\cdot{\small \bm{y}_q})} 
                       |\sigma_i \sigma_j \sigma_k\rangle   \rangle=  
} \label{eq38b} \\ & &
\sum_{s_{x_q}} 
\langle \Psi_1^{JM}|V_q(x_q) | 
  e^{i({\small \bm{k}}_{x_q}\cdot{\small \bm{x}_q}+{\small \bm{k}}_{y_q}\cdot{\small \bm{y}_q})} 
            | \chi_{s_{x_q} s_{y_q}}^{JM}  \rangle \rangle 
\langle  \chi_{s_{x_q} s_{y_q}}^{JM}  |\sigma_i \sigma_j \sigma_k\rangle,
\nonumber
\end{eqnarray}
where $s_{x_q}$ is the spin of the two-body system connected by the $\bm{x}_q$ Jacobi coordinate,
we can then, by use of Eqs.(\ref{eq38}) and (\ref{eq38b}), write the transition amplitude
in Eq.(\ref{eq37}) exactly as given in Eq.(\ref{eq25}), where now 
\begin{eqnarray}
\lefteqn{ \hspace*{-5mm}
A_{s_{x_q}}(q)=
\frac{\pi}{\sqrt{2}}
\frac{\kappa^2}{\sqrt{\kappa k_{y_q}}} \frac{1}{(2\pi)^3}
\left( \frac{\mu_x}{m} \right)^{3/4} e^{i \frac{\pi}{4}} 
} \label{eq41} \\ & &
\times \frac{2m}{\hbar^2} \langle \Psi_1^{JM}|V_q(x_q) | 
  e^{i({\small \bm{k}}_{x_q}\cdot{\small \bm{x}_q}+{\small \bm{k}}_{y_q}\cdot{\small \bm{y}_q})} 
            | \chi_{s_{x_q} s_{y_q}}^J  \rangle \rangle.
\nonumber
\end{eqnarray}

Thus, Eq.(\ref{eq41}) permits to obtain Eq.(\ref{eq27}), and
therefore the cross sections in Eqs.(\ref{eq9}) and (\ref{eq12}). Contrary to what happens in 
Eq.(\ref{eq26}), Eq.(\ref{eq41}) does not contain any expansion of the outgoing wave, and in principle 
the infinitely many breakup adiabatic terms are included. 
The discussion of the calculation of the matrix element contained in Eq.(\ref{eq41})
is given in the next section.

\section{Calculation of the breakup integral relation using the HA formalism}

Let us denote the matrix element to be computed as:
\begin{equation}
M_{s_{x_i}}(\bm{k}_{x_i},\bm{k}_{y_i})=\langle \Psi_1^{JM}|V_i(x_i) |  
  e^{i({\small \bm{k}}_{x_i}\cdot{\small \bm{x}_i}+{\small \bm{k}}_{y_i}\cdot{\small \bm{y}_i})} 
            | \chi_{s_{x_i} s_{y_i}}^{JM}  \rangle \rangle. 
\label{eq43}
\end{equation}

The three-body wave function $\Psi_1^{JM}$ is expanded in terms of the adiabatic angular 
eigenfunctions, which in turn are expanded in terms of the hyperspherical harmonics 
(see also Refs.~\cite{rom11,gar12}):
\begin{eqnarray}
\lefteqn{
\Psi_1^{JM}(\bm{x}_i,\bm{y}_i)=}  \label{expa}\\ & &
\frac{1}{\rho^{5/2}} \sum_n f_{n1}(\rho) \sum_{[K]} C^{(n)}_{[K]}(\rho)
{\cal Y}_{[K]}^{JM}(\Omega_i) \nonumber
\end{eqnarray}
where the coefficients $C^{(n)}_{[K]}$ reduce to $C^{(n)}_{K,s_{x_i}}$ in the
case of $s$-waves and
where we have selected the $i$ arrangement of the Jacobi coordinates to construct the
three-body HH-spin functions ${\cal Y}_{[K]}^{JM}$. It should be noticed that 
it is convenient to expand the
three-body wave function $\Psi_1^{JM}$ in the set $i$ of Jacobi coordinates in which
the coordinate $\bm{x}_i$ is the same appearing in the interaction potential, as
given in Eq.(\ref{eq43}). Therefore for each of the three integrals given in
Eq.(\ref{eq38}), the corresponding set of Jacobi coordinates is used.

Inserting the above expansion into Eq.(\ref{eq43}), we get that the matrix element takes the form 
(the index $i$ will be omitted from now on and we consider $s$-wave only):
\begin{widetext}
\begin{equation}
M_{s_x}(\bm{k}_x,\bm{k}_y)= \frac{1}{4\pi} 
\int d^3x d^3y \frac{1}{\rho^{5/2}} \sum_n f_{n1}(\rho) \sum_{K} C^{(n)}_{K,s_x}(\rho)
N_K P_\nu^{(\frac{1}{2},\frac{1}{2})}(\cos2\alpha)
\langle \chi_{s_{x} s_{y}}^{JM}  | V(x) |  \chi_{s_{x} s_{y}}^{JM}  \rangle
e^{i({\small \bm{k}}_{x_i}\cdot{\small \bm{x}_i}+{\small \bm{k}}_{y_i}\cdot{\small \bm{y}_i})}
\label{eq45}
\end{equation}
\end{widetext}
where the two-body potential is assumed not to mix different $s_x$-values.

Keeping in mind that the input in our calculation will be the directions of the momenta of two of the
outgoing particles and that, as already shown, for each value of the arclength $S$ it is possible to 
construct the full momenta of the three particles after the breakup, these three momenta permit us 
to construct $\bm{k}_x$ and $\bm{k}_y$, and the matrix element in Eq.(\ref{eq45}) can then be 
computed as a function of the arclength $S$.

Note also that in Eq.(\ref{eq45}), since only $s$-waves are assumed to contribute, the full dependence 
of the integrand on $\Omega_x$ and $\Omega_y$ is contained in the exponential. 
The integration over $\Omega_x$ and $\Omega_y$ can then be made analytically, leading to the
expression:
\begin{eqnarray}
\lefteqn{ \hspace*{-2cm}
\int d\Omega_x d\Omega_y 
   (\cos({\small \bm{k}}_{x}\cdot{\small \bm{x}}+{\small \bm{k}}_{y}\cdot{\small \bm{y}}) 
+ i \sin({\small \bm{k}}_{x}\cdot{\small \bm{x}}+{\small \bm{k}}_{y}\cdot{\small \bm{y}}))
= } \nonumber \\ & &
(2\pi)^2 \frac{4 \sin(k_x x)\sin(k_y y)}{k_x x k_y y},
\end{eqnarray}
which is real (the integral involving the sinus is just zero).

The remaining integral over $x$ and $y$ has to be performed numerically:
\begin{widetext}
\begin{eqnarray}
M_{s_x}(k_x,k_y)= 4\pi
\int dx dy \frac{x y}{\rho^{5/2}} \sum_n f_{n1}(\rho) \sum_{K} C^{(n)}_{K,s_x}(\rho)
N_KP_\nu^{(\frac{1}{2},\frac{1}{2})}(\cos2\alpha)
\langle \chi_{s_{x} s_{y}}^{JM}  | V(x) |  \chi_{s_{x} s_{y}}^{JM}  \rangle
\frac{\sin(k_x x)\sin(k_y y)}{k_x k_y}.
\label{eq47}
\end{eqnarray}
\end{widetext}

As we can see, the dependence on the directions of $\bm{k}_x$ and ${\bm k_y}$ has
disappeared. The dependence on the momenta is only through $k_x=\kappa \sin\alpha_\kappa$ and 
$k_y=\kappa\cos\alpha_\kappa$, where $\kappa=\sqrt{2mE/\hbar^2}$ and where $E$ is the three-body
energy above threshold. It is important not to get confused with $\alpha$ ($=\arctan x/y$), which
is a variable in coordinate space to be integrated away, and $\alpha_\kappa$ ($=\arctan k_x/k_y$),
which is a variable in momentum space, and which takes a well defined value for each value of
the arclength $S$. 

\begin{figure}
\centerline{\psfig{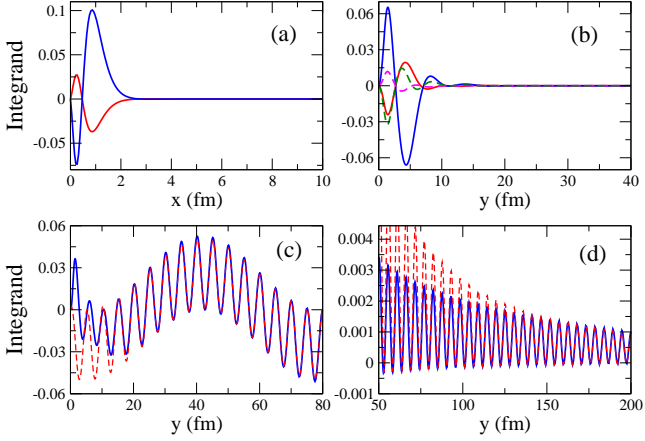}}
\caption{(Color online) Typical integrand of Eq.(\ref{eq47}) obtained for the $n-d$ reaction and for 
the following
different cases: (a) as a function of $x$ for two fixed values of $y$, (b) as a function of $y$ 
($x$-coordinate integrated away) and $n$ being a 1+2 channel for the neutron-neutron potential 
(solid curves), and for the neutron-proton
potential with $s_x=0$ (dashed curves), (c) as a function of $y$ ($x$-coordinate integrated away) 
and $n$ being a 1+2 channel for the neutron-proton potential and $s_x=1$, (d) as a
function of $y$ ($x$-coordinate integrated away) for $n$ being a breakup channel. In (c) and
(d) the dashed curves are the asymptotic matching given by Eq.(\ref{eq50}) and (\ref{eq55}), 
respectively.}
\label{fig2}
\end{figure}

The integral over $x$ is limited by the short-range character of the potential $V(x)$. 
Therefore the numerical computation of the integral does not present any particular 
difficulties. In Fig.~\ref{fig2}a 
we show two examples of the typical behavior of the integrand as a function of $x$ 
for two arbitrary fixed values of $y$. 
The curves correspond to the $n-d$ breakup reaction whose details will be given later on. They have been 
obtained for two fixed values of the $y$-coordinate. As we can see, the function becomes basically 
zero for rather small values of $x$.

The integral over $y$ is however more complicated. In general, the integrand does not fall off 
exponentially at large distances, and its calculation is therefore more delicate. 
We can distinguish two
different cases depending on the asymptotic behavior of the $f_{n1}(\rho)$ radial wave function, i.e.,
when $n$ is associated to the 1+2 channel or 
to a breakup channel. In the following we discuss the two cases.

\subsubsection{Index $n$ corresponding to a 1+2 channel}
\label{int1}

When $n$ labels a 1+2 channel ($n=1$ in our case), the corresponding Faddeev
amplitudes of the angular function $\Phi_n^{JM}(\rho,\Omega_\rho)$ behave 
at large distances as given by Eq.(\ref{apd1}), which reduces to
\begin{equation}
\Phi_n^{JM}(\rho,\Omega_i) \rightarrow \rho^{3/2}\frac{1}{4\pi} \Psi_d
                     | \chi_{s_{d} s_{y}}^{JM} \rangle,
\label{eq48}
\end{equation}
when only $s$-waves are involved. In this expression $\Psi_d$ is the bound two-body wave function of the 
dimer associated to the 1+2 channel $n$ and $s_d$ is the spin of the dimer.

Having this in mind, 
we can observe that the sum on the HH index $K$ in Eq.(\ref{eq47}) can be reconstructed
for $n=1$ (1+2 channel) as
\begin{eqnarray}
&\frac{1}{4\pi}\sum_{K} C^{(n)}_{K,s_x}(\rho)
N_K P_\nu^{(\frac{1}{2},\frac{1}{2})}(\cos2\alpha)
\langle \chi_{s_{x} s_{y}}^{JM}  | V(x) |  \chi_{s_{x} s_{y}}^{JM}  \rangle
=  \nonumber \\ 
&\langle \Phi_1^{JM}(\rho,\Omega_\rho)  | V(x) |  \chi_{s_{x} s_{y}}^{JM}  \rangle= \nonumber
\\
&\sum_i \langle \Phi_1^{JM}(\rho,\Omega_i)  | V(x) |  \chi_{s_{x} s_{y}}^{JM}  \rangle .
\end{eqnarray} 
The last expression shows
that, at large distances, the integrand in Eq.(\ref{eq47}) contains explicitly the
dimer wave function in the different Jacobi permutations. Two different possibilities
appear: the potential and the dimer depend on Jacobi coordinates belonging to two
different permutations, for example $\bm{x}_i$ and $\bm{x}_j$. In this case
the integral in
Eq.(\ref{eq47}) has an exponential fall off in both coordinates, $x$ and $y$.
In fact, when the two-body potential refers to the two particles that do
not form the dimer (for instance the $V_{nn}(x)$ potential in the $n-d$ case),
large values of $y$ would correspond to large distances between the third particle and
the one with which that third particle forms the dimer 
(large neutron-proton distance in the $n-d$ case).
Thus, for sufficiently large values of $y$ the integrand should be zero due to the presence
of the bound dimer wave function, which is now basically associated 
to the $y$-coordinates at large distances. Taking again the $n-d$ case as an illustration, the solid 
curves in Fig.\ref{fig2}b correspond to the integrand of Eq.(\ref{eq47}) as a function of $y$ 
for two different values of the arclength $S$
(after integrating away the $x$-coordinate) for the case of the
neutron-neutron potential. As shown in the figure, the integrand in the $y$ coordinate
dies pretty fast, and it is negligible already at distances of even less than 40 fm.

A similar case arises when the interaction is the one giving rise to the dimer 
(the neutron-proton potential in the $n-d$ case) and $s_x$ is different to the spin 
of the bound dimer ($s_x=0$ in the $n-d$ case). The reason now is that the coefficients
$C_{K,s_x\neq s_d}(\rho)$ in Eq.(\ref{eq47}) go to zero for large values of the hyperradius
(as shown in Eq.(\ref{eq48}), asymptotically only the terms with $s_x=s_d$ survive).
This case is illustrated in Fig.\ref{fig2}b by the dashed lines, which correspond to the
$n-d$ case, and which show that
also in this case the integrand goes to zero rather fast.
Therefore, in the particular cases shown in Fig.\ref{fig2}b the numerical computation of the integral
in the both variables, $x$ and $y$, do not present particular problems.

A different situation appears when the potential in Eq.(\ref{eq47}) and the dimer depend on the
same variable $\bm{x}$ and when
$s_x=s_d$ (neutron-proton potential and $s_x=1$ in the $n-d$ case).  In this case the
the $y$-coordinate integrand in Eq.(\ref{eq47}) does not vanish exponentially and
a particular analysis of the integrand tail has to be performed.

It is well known \cite{gar12} that the large distance behavior of $f_{n1}(\rho)$, for $n$ being 
a 1+2 channel, is given by:
\begin{equation}
f_{n1}(\rho) \rightarrow C \sin(k_y^{(n)}y) + D \cos(k_y^{(n)}y),
\label{eq50}
\end{equation} 
where $C$ and $D$ are complex numbers (the $f_{n1}$-functions are complex), and
$(k_y^{(n)})^2=2m(E-E_n)/\hbar^2$, where $E_n$ is the binding energy of the dimer
present in the 1+2 channel associated to the adiabatic term $n$. 

Therefore, using Eqs.(\ref{eq50}) and (\ref{eq48}) it is clear that the integrand in
Eq.(\ref{eq47}) is asymptotically separable into $x$ and $y$ coordinates, and the integrand of the
$y$ part goes like:
\begin{eqnarray}
\lefteqn{ \hspace*{-5mm}
\mbox{Integrand}(y) \rightarrow } \label{eq51} \\ & &
C \sin(k_y^{(n)}y)\sin(k_y y) + D \cos(k_y^{(n)}y) \sin(k_y y),
\nonumber
\end{eqnarray} 
where $C$ and $D$ are complex constants, and where we have used that, since $x$ is restricted
to small values, then $y/\rho \rightarrow 1$.

This behavior is shown in Fig.\ref{fig2}c again for the $n-d$ case. The solid curve is the integrand 
in the $y$-coordinate obtained numerically for some value of the arclength $S$, and the dashed curve 
is the matching with the expression in Eq.(\ref{eq51}). As shown
in the figure the matching is pretty good for distances already of about 30-40 fm.

Therefore an easy way of computing the integral over $y$ in Eq.(\ref{eq47}) (for the two-body potential giving
rise to the dimer and $s_x=s_d$) is to do it numerically after subtracting 
the asymptotic behavior given in Eq.(\ref{eq51}), i.e., integrating numerically the difference between 
the solid and the dashed curves in Fig.\ref{fig2}c, which dies asymptotically sufficiently fast, and 
afterward add the analytical integral of Eq.(\ref{eq51}) from $y=0$ to $\infty$.
This analytical integral can be easily made by using that:
\begin{widetext}
\begin{equation}
\int_0^\infty \sin(k_y^{(n)}y)\sin(k_y y) dy = 
\lim_{a\rightarrow 0}\left[ \int_0^\infty e^{-a y} \sin(k_y^{(n)}y)\sin(k_y y) dy \right] = 0 ; (a>0)
\label{an1}
\end{equation}
\begin{equation}
\int_0^\infty \cos(k_y^{(n)}y)\sin(k_y y) dy = 
\lim_{a\rightarrow 0}\left[ \int_0^\infty e^{-a y} \cos(k_y^{(n)}y)\sin(k_y y) dy \right] = 
  \frac{k_y}{(k_y)^2-(k_y^{(n)})^2} ; (a>0)
\label{an2}
\end{equation}
\end{widetext}
where we have to remember that $k_y=\kappa \cos\alpha_\kappa$ and
$k_y^{(n)}=\sqrt{2m(E-E_n)/\hbar^2}$, which are always different ($E_n<0$).

\subsubsection{Index $n$ corresponding to a breakup channel}
\label{int2}

Due to the short-range character of the potential, the $x$-values are in any case restricted 
to relatively small values, no matter the character of the channel $n$ in Eq.(\ref{eq47}).
For this reason, even if $n$ corresponds to a breakup channel,
the large distance behavior of the integrand is given by contributions
fulfilling that $ y\gg x$ (or, in other words, $\rho \approx y$).

Also, for a breakup channel $n$, the coefficients $C_{K,s_x}^{(n)}(\rho)$ go to constant values
(the angular eigenfunctions $\Phi_n^J$ become just a linear combination of hyperspherical harmonics,
see Eq.(\ref{eq21})),
and the Jacobi Polynomials $P_\nu^{(\frac{1}{2},\frac{1}{2})}(\cos 2\alpha)$ contained in
the hyperspherical harmonics go also to the constant value $P_\nu^{(\frac{1}{2},\frac{1}{2})}(1)$.

Finally, the general behavior at large distances of the radial functions $f_{n1}(\rho)$ is given 
by a linear combination of the Hankel functions of first and second order \cite{gar12}, which means
that their asymptotic behavior is given by:
\begin{equation}
f_{n1}(\rho) \rightarrow \sum_{m=0}^M  \frac{C_m\sin(\kappa y) + D_m \cos(\kappa y)}{y^m},
\label{eq54}
\end{equation}
where we have already replaced $\rho$ by $y$.

Therefore, when $n$ is associated to a breakup channel, the $y$-part of the integrand in
Eq.(\ref{eq47}) goes asymptotically as:
\begin{eqnarray}
\lefteqn{
\mbox{Integrand}(y) \rightarrow }
   \label{eq55} \\ && 
    \sum_{m=0}^M
\left( \frac{C_m \sin(\kappa y)\sin(k_y y)}{y^{m+3/2}} + \frac{D_m \cos(\kappa y) \sin(k_y y)}{y^{m+3/2}} \right),
\nonumber
\end{eqnarray}
where, again, the constants $C_m$ and $D_m$ are complex and $k_y = \kappa \cos \alpha_\kappa$. Obviously, 
the higher the value of $M$, the lower the value of $y$ at which the matching with the numerical integrand
is obtained.

Thus, for outgoing breakup channels, the $y$-part of the integral in Eq.(\ref{eq47})
goes to zero asymptotically as $1/y^{3/2}$. It could then seem that such integral could be done numerically
without much trouble. However, the fall off is not fast enough, and at relatively large distances the
integrand is not really negligible. Taking again the $n-d$ case as an example, this is shown in 
Fig.\ref{fig2}d, where the blue curve is the integrand in Eq.(\ref{eq47}) as a function of $y$
for some value of the arclength $S$ and
for $n$ being one of the breakup channels ($n>1$). At a distance of about 200 fm the integrand is
not at all negligible, in fact the amplitude of the oscillations shown in the figure are still
about 10\% of the maximum computed amplitude. The dashed curve shows the matching given 
by Eq.(\ref{eq55}) and $M=0$. It is clear
that this matching is less accurate than the one shown in Fig.\ref{fig2}c, and the agreement
now with the true integrand is observed much farther, at 150 fm at least.

Therefore, due to the too slow fall off of the integrand, it is convenient to integrate numerically 
up to some $y_{\mbox{\scriptsize max}}$ value at which the matching with the asymptotic behavior 
in Eq.(\ref{eq55}) has already been achieved, and to perform analytically 
the integrand from $y_{\mbox{\scriptsize max}}$ up to $\infty$. These analytical integrations
involve the so-called Fresnel integrals, and they can be performed as indicated in appendix \ref{app6}.

\section{The $n-d$ case}

In this section we shall compute the cross sections for neutron-deuteron breakup. This choice is made
in order to compare to the benchmark calculations given in \cite{fri95}. Following this reference, we 
have chosen the Malfliet-Tjon-I-III model $s$-wave nucleon-nucleon potential \cite{fri90}, which for the
triplet and singlet cases takes the form:
\begin{equation}
V_t(r)=\frac{1}{r} \left(
-626.885 e^{-1.55 r} + 1438.72 e^{-3.11 r}
\right),
\label{eq60}
\end{equation}
and
\begin{equation}
V_s(r)=\frac{1}{r} \left(
-513.968 e^{-1.55 r} + 1438.72 e^{-3.11 r}
\right),
\label{eq61}
\end{equation}
respectively, and where the $r$ is in fm, the potential in MeV, and $\hbar^2/m=41.47$ MeV fm$^2$.
In our calculations $m$ is taken as the normalization mass in the definitions of the Jacobi coordinates 
given in Eqs.(\ref{eq1}) to (\ref{eq4}).

In the first part of this section we summarize the results obtained in Ref.\cite{gar12} for 
the ${\cal S}$-matrix after application of the integral relations. We continue with the cross sections 
obtained from the adiabatic expansion of the transition amplitude in Eq.(\ref{eq26}), and the inaccuracies 
corresponding to some particular geometries are analyzed. In the last part of the section we show how the use
of the transition amplitude given by Eq.(\ref{eq41}) corrects the cross sections in the regions were 
the inaccuracies were observed.

\subsection{${\cal S}$-matrix}

The details concerning the integral relations formalism applied to the description of breakup 1+2
reactions are given in Ref.\cite{gar12}. This method permits to extract the ${\cal S}$-matrix of the
reaction from the internal part of the wave function. When this wave function is obtained by means of
the adiabatic expansion method, the pattern of convergence of the ${\cal S}$-matrix is similar to the one
obtained with the same method for bound states.

In \cite{gar12} the integral relations method is applied to the $n-d$ reaction.
Only $s$-waves are considered in the calculation, which implies that only
two different total angular momenta are possible, the quartet
case ($J$=3/2), for which only the triplet $s$-wave potential
in Eq.(\ref{eq60}) enters, and the doublet case ($J$=1/2), for which both
the singlet and the triplet potentials contribute.

The unitarity of the ${\cal S}$- matrix implies that given an incoming
channel, for instance, channel 1 ($n-d$ channel), we have that 
$\sum_{n=1}^\infty |{\cal S}_{1n} |^2=1$, or in other words:
\begin{equation}
\sum_{n=2}^\infty |{\cal S}_{1n} |^2=1-|{\cal S}_{11} |^2,
\end{equation}
which means that an accurate calculation of the elastic term ${\cal S}_{11}$ 
amounts to an accurate calculation of the infinite summation of
the $|{\cal S}_{1n} |^2$ terms ($n > 1$) corresponding to the breakup channels, 
and therefore allowing the calculation of the total breakup cross section 
given in Eq.(\ref{crossb}).

The complex value of ${\cal S}_{11}$ can be written in terms of a
complex phase shift $\delta$ as
\begin{equation}
{\cal S}_{11}=e^{2i\delta}=e^{-2 \mbox{\scriptsize Im}(\delta)} e^{2i \mbox{Re}(\delta)}
=|{\cal S}_{11}|e^{2i \mbox{Re}(\delta)}.
\label{eq63}
\end{equation}

The value of $|{\cal S}_{11}|^2$ gives the probability of elastic neutron-deuteron scattering, and
$|{\cal S}_{11}|$ is what usually referred to as the inelasticity parameter (denoted by $\eta$ in
\cite{fri95,fri90}).  Obviously, the closer the inelasticity to 1 the more elastic the reaction.
In fact, for energies below the breakup threshold the phase-shift is real and $|{\cal S}_{11}|=1$.

\begin{table}
\caption{Inelasticity parameter $|{\cal S}_{11}|$ for the neutron-deuteron scattering for two
different laboratory neutron beam energies (14.1 MeV and 42.0 MeV) for the doublet and
quartet cases. The value of $K_{\mbox{\scriptsize max}}$ is the $K$-value associated 
to the last adiabatic potential included in the calculation. 
The last row gives the value quoted in Ref.\cite{fri95}.}
\label{tab1}
\begin{tabular}{c|cc|cc}
  $K_{\mbox{\scriptsize max}}$   & \multicolumn{2}{c|}{Doublet} & \multicolumn{2}{c}{Quartet} \\ \hline
       &      14.1 MeV         &       42.0 MeV       &       14.1  MeV        &     42.0  MeV          \\ \hline
   4   &    0.4662       &     0.4929     &     0.9794       &   0.8975         \\
   8   &    0.4637       &     0.4993     &     0.9784       &   0.9026         \\
  12   &    0.4640       &     0.5014     &     0.9783       &   0.9030         \\
  16   &    0.4643       &     0.5019     &     0.9782       &   0.9031         \\
  20   &    0.4644       &     0.5021     &     0.9782       &   0.9033         \\
  24   &    0.4645       &     0.5022     &     0.9782       &   0.9033         \\
  28   &    0.4645       &     0.5022     &     0.9782       &   0.9033         \\
                                                                                     \hline
Ref.\cite{fri95} &    0.4649    &  0.5022    &   0.9782      &   0.9033   \\
\end{tabular}
\end{table}

\begin{table}
\caption{The same as Table~\ref{tab1} for Re($\delta$).}
\label{tab2}
\begin{tabular}{c|cc|cc}
  $K_{\mbox{\scriptsize max}}$   & \multicolumn{2}{c|}{Doublet} & \multicolumn{2}{c}{Quartet} \\ \hline
        &         14.1 MeV     &      42.0 MeV      &      14.1  MeV      &    42.0  MeV         \\ \hline
   4    &      105.82    &    42.66     &    69.04      &   38.98        \\
   8    &      105.57    &    41.65     &    68.99      &   37.95        \\
  12    &      105.53    &    41.49     &    68.98      &   37.77        \\
  16    &      105.53    &    41.46     &    68.97      &   37.73        \\
  20    &      105.53    &    41.45     &    68.96      &   37.72        \\
  24    &      105.53    &    41.44     &    68.96      &   37.71        \\
  28    &      105.53    &    41.44     &    68.96      &   37.71        \\
                                                                                                   \hline
Ref.\cite{fri95}   &  105.50    &   41.37     &   68.96       &    37.71   \\
\end{tabular}
\end{table}

In tables~\ref{tab1} and \ref{tab2} we quote the results given in \cite{gar12} for the inelasticity 
parameter $|{\cal S}_{11}|$ and the real part of the phase shift Re($\delta$), respectively.
We have used the same two lab energies considered in Ref.\cite{fri95}, i.e., 14.1 MeV and 42.0 MeV.
The value of $K_{\mbox{\scriptsize max}}$ given in the tables corresponds to the asymptotic grand-angular 
quantum number associated to the last adiabatic potential included in the adiabatic expansion.

As seen in the tables, the agreement with the results in Ref.\cite{fri95} is good. Actually, we obtain 
precisely the same result for the two incident energies in the quartet case,
and a small difference clearly smaller than 0.1\% in the doublet case. Furthermore, the pattern of
convergence is rather fast, specially in the quartet case, for which already for
$K_{\mbox{\scriptsize max}}=8$ we obtain a result that can be considered very accurate. In the
doublet case the convergence is a bit slower, and a value of $K_{\mbox{\scriptsize max}}$ of about
16 is needed. 

\subsection{Cross sections}

Once the ${\cal S}$-matrix of the reaction has been computed, we can now obtain the transition matrix
according to the expression in Eq.(\ref{eq26}), and therefore also the expression Eq.(\ref{eq27}) and the 
lab cross section in Eq.(\ref{eq12}).

We shall consider an incident energy in the lab frame of $E_{in}^{(lab)}=14.1$ MeV and four different 
outgoing geometries specified by the polar angles $\theta_1$ and 
$\theta_2$ and the difference of azimuthal angles $\Delta \varphi=\varphi_1-\varphi_2$, where 
$(\theta_1,\varphi_1)$ and $(\theta_2,\varphi_2)$ describe the direction of the two outgoing neutrons.
The four different cases are:
\begin{itemize}
\item Case 1: $\theta_1=45.0^\circ$, $\theta_2=53.56^\circ$, $\Delta \varphi=180^\circ$.
\item Case 2: $\theta_1=35.0^\circ$, $\theta_2=44.0^\circ$, $\Delta \varphi=180^\circ$.
\item Case 3: $\theta_1=60.06^\circ$, $\theta_2=53.10^\circ$, $\Delta \varphi=180^\circ$.
\item Case 4: $\theta_1=51.02^\circ$, $\theta_2=51.02^\circ$, $\Delta \varphi=120^\circ$.
\end{itemize}

\begin{figure}
\centerline{\psfig{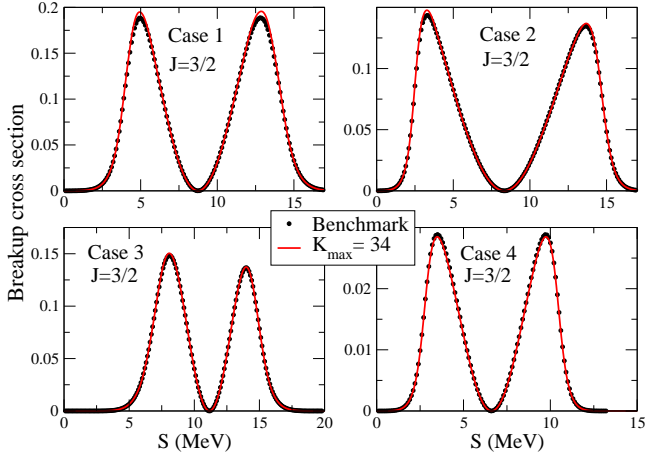}}
\caption{(Color online) Breakup cross sections, Eq.(\ref{eq12}), as a function of the arclength $S$
for $n-d$ scattering in the quartet case ($J=3/2$) for a lab incident energy of 14.1 MeV. The 
cross sections are given in mb/(MeV sr$^2$). The 
value of $K_{max}$ refers to the grand-angular quantum number $K$
associated to the last adiabatic term included in the expansion in Eq.(\ref{eq26}). 
The four cases shown in the figure correspond to the four 
different directions of the outgoing neutrons specified in the text. 
The dotted curves are the result
of the benchmark calculations given in Ref.\cite{fri95}.}
\label{fig3}
\end{figure}

The solid lines in Fig.\ref{fig3} show the computed cross section in Eq.(\ref{eq12}) for the four 
cases given above for the quartet case ($J=3/2$). In the figure, $K_{max}$ refers to 
the asymptotic value of the grand-angular quantum number $K$ associated to last adiabatic 
term included in the expansion in Eq.(\ref{eq26}). A value of $K_{max}=34$ amounts to 
including 18 adiabatic terms in the expansion. This is enough to reach convergence. 
In fact, the same calculation with $K_{max}=22$ 
(12 adiabatic terms) is basically indistinguishable from the curves shown in the figure. 
The cross sections given in the benchmark calculation in Ref.\cite{fri95} are shown by 
the dotted curves. As we can see, the results given in the benchmark calculation are 
nicely reproduced.

\begin{figure}
\centerline{\psfig{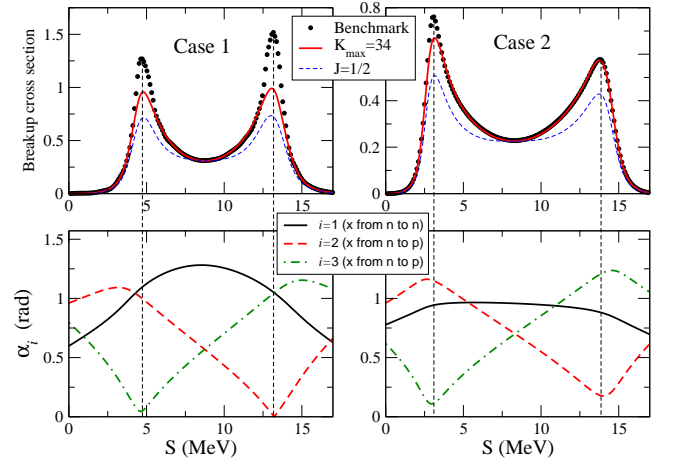}}
\caption{(Color online) 
Upper part: Total cross sections for the Cases 1 and 2 in the upper part of Fig.\ref{fig3}. They are
obtained by adding the quartet ($J=3/2$) contribution in Fig.\ref{fig3} and the doublet ($J=1/2$)
contribution shown by the dashed curves. Lower part: For the same two cases, the corresponding values of 
$\alpha_i$ ($i=1,2,3$) as a function of the arclength $S$.}
\label{fig4}
\end{figure}

\begin{figure}
\centerline{\psfig{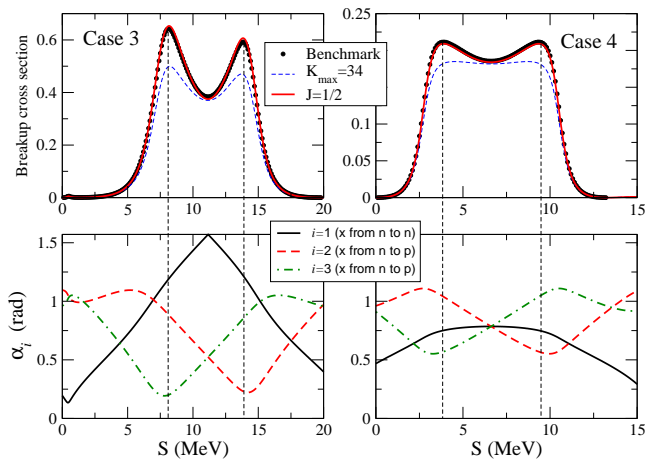}}
\caption{(Color online) The same as in Fig.\ref{fig4} for the Cases 3 and 4.}
\label{fig5}
\end{figure}

The corresponding cross sections for the doublet case ($J=1/2$) are shown by the 
dashed curves in the upper parts of Figs.\ref{fig4} and \ref{fig5}.
When the contribution from the quartet state (Fig.\ref{fig3}) is added, we get the 
total cross section given by the
solid curves in the same figures. The doublet cross sections have been obtained with $K_{max}=34$, 
which for $J=1/2$ corresponds to inclusion of 35 adiabatic terms in the expansion of Eq.(\ref{eq26}). 
Again, when
$K_{max}$ is reduced to 22 (23 adiabatic terms) the computed curves can not be distinguished from 
the ones shown in the figure. As we can see, even if the computed cross sections have converged, there 
is a clear discrepancy with the benchmark calculation for Case 1, and, to a lower extent, for 
Case 2 (Fig.\ref{fig4}). For Cases 3 and 4 (Fig.\ref{fig5}) the agreement is reasonably good.

As anticipated in Section~\ref{2d}, the discrepancy with the benchmark calculation
appears in the $S$-regions corresponding to an outgoing kinematics where two of the particles 
have zero relative energy. This can be seen in Fig.\ref{fig4}, where the lower part shows, 
as a function of the 
arclength $S$, the hyperangle $\alpha_i=\arctan(k_{x_i}/k_{y_i})$ for the Cases 1 and 2. 
The index $i=1$ corresponds to the Jacobi coordinates where $\bm{x}$ connects the two neutrons
(solid curve), and the indices $i=2,3$ correspond to the Jacobi coordinates where $\bm{x}$ connects 
the proton and one of the neutrons (dashed and dot-dashed curves).
For Case 1 (left part of the figure), we can see that $\alpha_2$ and $\alpha_3$ approach zero 
very much for $S$ slightly below 5 MeV and about 13 MeV. In these regions the proton and one of
the neutrons move with relative zero (or very small) energy, and these are precisely the regions
where the discrepancy with the benchmark calculation is observed.
For Case 2, $\alpha_{2}$  and $\alpha_{3}$ do not approach zero as much as in Case 1, and 
the discrepancy with the benchmark calculation is now much smaller than in Case 1 (or even not visible,
as it happens for the second peak of the cross section).

For completeness, we show in the lower part of Fig.\ref{fig5} the same as in Fig.\ref{fig4} but 
for Cases 3 and 4. Now the
values of $\alpha_i$ do not reach zero for any $S$-value, and the agreement between our calculation and
the benchmark is fine for all values of $S$.

As already mentioned, the problem is related to the fact that $\alpha_i\approx 0$ means two of the particles
flying together with relative zero (or very small) energy, and the third particle moving far apart. The
adiabatic expansion in Eq.(\ref{eq26}) is then trying to reproduce the zero energy two-body wave function
(which is proportional to $\sin \alpha$) in terms of a set of polynomials (the Jacobi polynomials) that
have $\cos 2\alpha$ as argument. A correct description of the zero energy two-body wave function would
require then infinitely many adiabatic terms.

As seen in Fig.\ref{fig3}, the inaccuracy mentioned above for $\alpha_i\approx 0$
is not visible in the quartet case, where only triplet nucleon-nucleon components enter. However, 
as seen in Fig.\ref{fig4}, where the contribution from the doublet has been added, the mismatch with 
the benchmark calculation becomes very significant. This is due to the fact that in the doublet case 
there is an important contribution from the singlet nucleon-nucleon components. This little difference
is actually very relevant, because the singlet nucleon-nucleon potential has a rather large scattering 
length of about $-23$ fm, more than four 
times bigger (in absolute value) than the scattering length of the triplet nucleon-nucleon
potential (of about $5$ fm).
This means that the $s$-wave singlet nucleon-nucleon system has a pretty low-lying virtual state 
that favors the structure mentioned above of two nucleons flying together after the collision at a 
very low relative energy. For the case of the triplet nucleon-nucleon state the energy of the 
corresponding virtual state is about 20 times higher than in the singlet case (the virtual state 
energy is proportional to the inverse of the scattering length squared).

The solution suggested in order to solve the disagreement observed in Fig.\ref{fig4}
 is to skip the expansion of the outgoing wave
in terms of the different adiabatic channels and compute instead the transition amplitude
as given in Eq.(\ref{eq41}), where the adiabatic expansion does not enter. The results obtained by
using this alternative method are given in the following section.

\subsection{Corrections to the cross sections}

The cross section in Eq.(\ref{eq13}) is now computed making use of Eq.(\ref{eq41}).
The matrix element in this expression is given by Eq.(\ref{eq47}), whose calculation requires 
using the techniques described is Sections~\ref{int1} and \ref{int2} when integrating over the 
$y$-coordinate. For the case described in Fig.\ref{fig2}c the integral over
$y$ is made by integrating numerically the difference between the computed integrand and its analytic
asymptotic behavior (difference between the solid and dashed curves in Figs.\ref{fig2}c),
and adding the analytical integral, from zero to $\infty$, of the asymptotic behavior. These 
analytical integrals can be made with the help of Eqs.(\ref{an1}) and (\ref{an2}). 
For the case described in Fig.\ref{fig2}d the integral over $y$ is made by integrating numerically
up to some $y_{\mbox{\scriptsize max}}$ at which the asymptotic behavior in Eq.(\ref{eq55}) has already
been reached, and the integral from $y_{\mbox{\scriptsize max}}$ to $\infty$ is made as described 
in appendix~\ref{app6}.

The integral in Eq.(\ref{eq47}) contains a summation over the adiabatic terms $n$ and the grand-angular
quantum number $K$.
This double summation comes from the expansion of the incident three-body wave function $\Psi_1$ given 
in Eq.(\ref{expa}). The convergence in terms of $n$ has been found to be fast, and use of around 20 
adiabatic terms is enough to get a converged result. However, the summation over $K$ is much more delicate. 
In fact, in the $S$-regions where the discrepancy between the old calculation and the benchmark was found, 
the number of hyperspherical harmonics needed in Eq.(\ref{eq47}) can be pretty high.

\begin{figure}
\centerline{\psfig{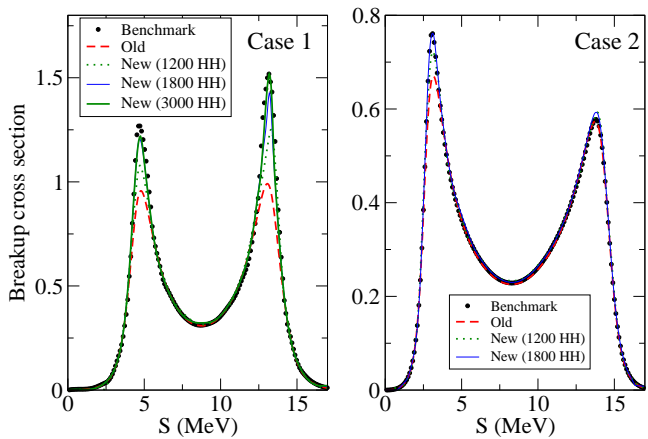}}
\caption{(Color online) Total cross sections (in mb/(MeV sr$^2$)) for the Cases 1 and 2 shown in
the upper part of Fig.\ref{fig4} 
($E_{in}^{(lab)}= 14.1$ MeV). The solid circles are the benchmark calculations in Ref.\cite{fri95}. 
The dashed curves are 
the results shown in Fig.\ref{fig4}, which were 
computed using the expansion in Eq.(\ref{eq26}). The cross sections obtained by means of 
Eq.(\ref{eq41}) are shown by the dotted, thin-solid, and thick-solid curves, which
correspond, respectively, to calculations including 1200, 1800, and 3000 
hyperspherical harmonics when computing the integral in Eq.(\ref{eq47}).}
\label{fig6}
\end{figure}

This is shown in Fig.\ref{fig6}, where we show the same cross sections as in Fig.\ref{fig4}
for the Cases 1 and 2. The benchmark result is given by the solid circles and the old computed cross 
sections, as shown in Fig.\ref{fig4}, are given by the dashed curves. The cross sections obtained with
the new procedure have been computed including 35 adiabatic terms in the expansion in Eq.(\ref{eq47}),
and the number of hyperspherical harmonics used in the same expansion are 1200 for the dotted curves,
1800 for the thin-solid curves, and 3000 for the thick-solid curve.

Needless to say, when the same procedure is used to compute the cross sections for the Cases 3 and 4, the
same results as the ones shown for these two cases in Fig.\ref{fig5} are obtained.

As we can see, in the regions where the results shown in Fig.\ref{fig3} were matching the benchmark, 
the new calculations still reproduce the results equally well. However, in the peaks where the 
old calculation and the benchmark did not agree, the new calculation improves the agreement
significantly. As already mentioned, the agreement is actually reached when a sufficiently high 
number of hyperspherical harmonics is included in the calculation. In fact, as we can see, 1200 
hyperspherical harmonics are still not enough, and 1800 of them are at least needed in Case 2, and
3000 in Case 1.

\begin{figure}
\centerline{\psfig{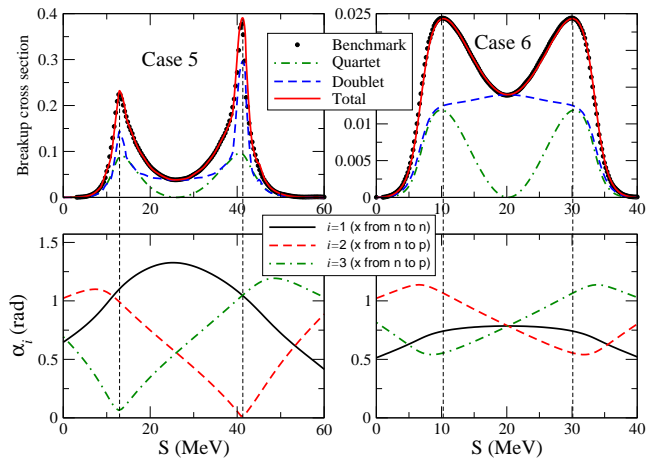}}
\caption{(Color online) Upper part: Total cross sections (in mb/(MeV sr$^2$)) for the Cases 5 and 6 and 
$E_{in}^{(lab)}=42.0$ MeV (see text). The solid circles are the benchmark calculations in 
Ref.\cite{fri95}. The dot-dashed and dashed curves are the quartet ($J=3/2$) and doublet ($J=1/2$)
contributions, respectively. The total cross sections are given by the solid curves. For Case 6 (right
panel) the calculation using the expansion in Eq.(\ref{eq26}) is accurate enough. For Case 5 (left panel)
the expansion in Eq.(\ref{eq41}) is required in order to obtain sufficient accuracy in the vicinity of the
peaks. Lower part: For the same two cases, the corresponding values of
$\alpha_i$ ($i=1,2,3$) as a function of the arclength $S$.}
\label{fig7}
\end{figure}

For completeness, we also show the results for the two benchmark calculations given in Ref.\cite{fri95}
corresponding to an incident energy in the laboratory frame of 42.0 MeV. The angles describing the 
direction of the outgoing neutrons in these two cases are:
\begin{itemize}
\item Case 5: $\theta_1=45.0^\circ$, $\theta_2=60.54^\circ$, $\Delta \varphi=180^\circ$.
\item Case 6: $\theta_1=53.61^\circ$, $\theta_2=53.61^\circ$, $\Delta \varphi=120^\circ$.
\end{itemize}

The results for Cases 5 and 6  ($E_{in}^{(lab)}=42.0$ MeV) are shown in Fig.\ref{fig7}.
As in Figs.\ref{fig4} and \ref{fig5}, the lower part shows the values of the hyperangle $\alpha_i$ 
as a function of the arclength $S$. As we can see, 
the Case 5 given above is analogous to the Cases 1 and 2 shown in Fig.\ref{fig4}, where 
the hyperangles $\alpha_2$ and $\alpha_3$ approach zero for certain values of the arclength $S$.
Now this happens for values of $S$ around 13 MeV and 40.5 MeV. For this reason, an
accurate calculation of the cross section for Case 5 in the vicinity of these two $S$-values
requires use of the expansion given in Eq.(\ref{eq41}). For Case 6 the value of the hyperangle
$\alpha$ is always far from zero (as in the Cases 3 and 4 shown in Fig.\ref{fig5}),
and use of the expansion in Eq.(\ref{eq26}) is enough to get a sufficient accuracy in the
calculations.

The upper part of Fig.\ref{fig7} shows the corresponding cross sections.
The dot-dashed and dashed curves
are the quartet ($J=3/2$) and doublet ($J=1/2$) contributions, respectively. The total cross section
is given by the solid curves. As we can see, the agreement with the benchmark calculations
(solid circles) given in Ref.\cite{fri95} is very good. In Case 5 the convergence in the vicinity
of the two peaks, computed through Eq.(\ref{eq41}), has required inclusion of up to 40 adiabatic
terms in the expansion of the three-body wave function (see Eq.(\ref{eq17})). This number is higher
than in the calculations for $E_{in}^{(lab)}=14.1$ MeV, where even less than 30 adiabatic channels
were enough. This is because for a given value of the hyperradius $\rho$ only
the adiabatic potentials whose value at $\rho$ is smaller than the incident energy (or close enough to the
energy if they are above) contribute to the radial wave functions $f_{n1}(\rho)$ obtained
through Eq.(\ref{radeq}). Obviously, the higher the three-body energy $E$ the higher the number
of adiabatic potentials contributing to the radial wave functions.

\section{Summary and conclusions}

The description of three-body systems, and in particular of $1+2$ reactions, by use of the adiabatic
expansion method has been proved to be efficient provided that the integral relations introduced in 
Ref.\cite{bar09b} are used. Because of these two relations the ${\cal S}$-matrix of the reaction
can be extracted from the internal part of the wave function, and subsequently the pattern of
convergence of the adiabatic expansion is fast. Knowledge of the ${\cal S}$-matrix permits then
to obtain accurate results for the total cross sections for the different open channels (elastic,
inelastic, transfer, or breakup).

However, even if the ${\cal S}$-matrix and the total cross sections can be accurately computed, 
it is not obvious that the differential cross sections that one extracts from the ${\cal S}$-matrix
are also accurate. In fact, the calculation of the differential cross sections requires knowledge of
not only of the ${\cal S}$-matrix, but of the full transition amplitude. When the adiabatic expansion
is applied to obtain the transition amplitude, the number of adiabatic terms necessary in 
order to achieve convergence strongly depends on the geometry of the outgoing particles, and under
some circumstances this number can be exceedingly large.

In this work we have given the details of how to compute differential cross sections for 1+2 reactions
above the threshold for breakup of the dimer. The expressions that permit to transform the cross 
sections from the center of mass frame to the laboratory frame have been derived. The three-body
wave function describing the three-body system as well as the transition amplitude are computed
by use of the adiabatic expansion method.

The case of neutron-deuteron breakup has been taken as a test, and our results have been compared
with the benchmark calculations described in Refs.\cite{fri95,fri90}. Six different geometries have been
considered for the three outgoing nucleons after the breakup. 

In general, the agreement with the benchmark results has been found to be good. The only discrepancy 
appears in those cases corresponding to two outgoing particles flying away together after the breakup 
with zero relative energy. In this case the adiabatic expansion of the transition amplitude is not 
convenient, since the number of adiabatic terms needed to reproduce such structure is too large.

We have shown that this problem can be solved by skipping the partial wave expansion of
the outgoing wave, and expressing the transition amplitude in terms of a full, non-expanded, 
outgoing plane wave. This calculation is much more demanding from the numerical point of view, 
and special care has to be taken with the number of hyperspherical harmonics used to describe 
the incoming wave function.

This study can be seen as a first step in the inclusion of the Coulomb interaction 
in the study of 1 + 2 nuclear processes. Preliminary studies using the HA 
expansion have been done in Ref.~\cite{kie10} for the elastic channel of the $p-d$
reaction. In the breakup case, Eq.(\ref{eq47}) can not be applied directly
since the free plane wave was used in its derivation, and it should be
replaced by a distorted wave due to the long range of the Coulomb potential.
This study is at present underway following the analysis given in
Ref.\cite{viv01}. However, it should be noticed that the configuration in which
two protons move away together is suppressed and the one in which a
neutron-proton pair moves away together far from the second proton
shows tiny Coulomb effects (see Ref.\cite{del05}). Hence, we expect to extend
the analysis here presented to the $p-d$ case, and in general to
$1+2$ nuclear processes allowing us to analyze several reactions of
astrophysical interest in which the inclusion of the Coulomb interaction cannot 
be avoided.

\acknowledgments
This work was partly supported by funds provided by DGI of MINECO (Spain) under 
contract No. FIS2011-23565.

\appendix

\section{The outgoing flux}
\label{app1}

The time-dependent Schr\"{o}dinger equation in hyperspherical coordinates is given by \cite{nie01}:
\begin{equation}
i\hbar \frac{\partial \Psi}{\partial t}=
\frac{\hbar^2}{2m} \left( 
-\frac{\partial^2 \Psi}{\partial \rho^2} -\frac{5}{\rho}\frac{\partial \Psi}{\partial \rho}
\right) + \frac{\hbar^2}{2m} \frac{\hat{\Lambda}}{\rho^2} \Psi+ \hat{V}\Psi,
\label{apen1}
\end{equation} 
where $\Psi$ is the three-body wave function, $\hat{\Lambda}$ is the grand-angular operator whose
eigenfunctions are the hyperspherical harmonics, and $\hat{V}$ is the operator containing the
two-body potentials.

The self-adjoint of the equation above is just:
\begin{equation}
-i\hbar \frac{\partial \Psi^\dag}{\partial t}=
\frac{\hbar^2}{2m} \left( 
-\frac{\partial^2 \Psi^\dag}{\partial \rho^2} -\frac{5}{\rho}\frac{\partial \Psi^\dag}{\partial \rho}
\right) + \frac{\hbar^2}{2m} \Psi^\dag \frac{\hat{\Lambda}}{\rho^2} + \Psi^\dag \hat{V},
\label{apen2}
\end{equation} 

Multiplying Eq.(\ref{apen1}) by $\Psi^\dag$ from the left and Eq.(\ref{apen2}) by $\Psi$ from the right
we get two expressions that after subtracting lead to:
\begin{equation}
\rho^5 \frac{\partial |\Psi|^2}{\partial t}= - \frac{1}{i}\frac{\hbar}{2m} \frac{\partial}{\partial \rho}
\left[
\Phi^\dag \frac{\partial \Phi}{\partial \rho} - \frac{\partial \Phi^\dag}{\partial \rho} \Phi
\right],
\label{apen3}
\end{equation}
where $\Phi = \rho^{5/2}\Psi$.

Therefore, the number of particles per unit time through the unit of surface of a hypersphere 
of hyperradius $\rho$ is given by: 
\begin{equation}
\frac{1}{i}\frac{\hbar}{2m} \frac{1}{\rho^5}\left[
\Phi^\dag \frac{\partial \Phi}{\partial \rho} - \frac{\partial \Phi^\dag}{\partial \rho} \Phi
                           \right].
\label{apen4}
\end{equation}

Thus, the flux of particles through an element of hypersurface $d\Sigma$ is given by:
\begin{equation}
\mbox{outgoing flux}=\frac{1}{i}\frac{\hbar}{2m} \frac{1}{\rho^5} \left[
\Phi^\dag \frac{\partial \Phi}{\partial \rho} - \frac{\partial \Phi^\dag}{\partial \rho} \Phi
                           \right] d\Sigma
\label{apen5}
\end{equation}

The asymptotic behavior of the three-body wave function in hyperspherical coordinates takes the form:
\begin{equation}
\Psi \stackrel{\rho \rightarrow \infty}{\longrightarrow}
\frac{1}{\rho^{5/2}} e^{i\kappa \rho} A,
\label{apen6}
\end{equation}
where $\kappa=\sqrt{2m E}/\hbar$, $E$ is the total three-body energy, and $A$ is the transition amplitude.

Keeping in mind that $\Phi=\rho^{5/2}\Psi$, substituting Eq.(\ref{apen6}) into Eq.(\ref{apen5}), and
using that the element of hypersurface is given by Eq.(\ref{eq6}), we get the following final expression
for the outgoing flux:
\begin{equation}
\mbox{outgoing flux}= \hbar \frac{\kappa}{m} |A|^2
\left(\frac{m}{\mu_x} \right)^{3/2} \left(\frac{m}{\mu_y} \right)^{3/2} d\Omega_\kappa,
\label{apen7}
\end{equation}
where we have used that, asymptotically, $d\Omega=d\Omega_\kappa$.

\section{Phase space for three particles with equal mass.}
\label{app2}

The choice of three particles with equal mass is made just to simplify the notation and the expressions 
derived in the following subsections. The mass of the particles $m$ is also taken as 
the normalization mass in the definition of the Jacobi coordinates given in Eqs.(\ref{eq1})
to (\ref{eq4}). In any case the extension of the expressions below to particles with different mass 
is straightforward.

\subsection{Center of mass frame}

To write down the phase space in the center of mass frame we use the momentum coordinates 
$\{\bm{p}_x,\bm{p}_y,\bm{P}\}$, where $\bm{p}_x$ and $\bm{p}_y$ are defined in Eqs.(\ref{eq3}) and
(\ref{eq4}), and $\bm{P}=\bm{p}_i+\bm{p}_j+\bm{p}_k$ is the momentum of the three-body center of mass. 
In these coordinates the phase space takes the form:
\begin{equation}
\Phi=\int d\bm{p}_x d\bm{p}_y d\bm{P} \delta(E-\frac{\hbar^2 p_x^2 }{2\mu_x}-\frac{\hbar^2 p_y^2 }{2\mu_y} ) 
     \delta^{(3)}(\bm{P}-\bm{p}_i-\bm{p}_j-\bm{p}_k),
\label{apb1}
\end{equation}
where the delta functions impose energy and momentum conservation ($E$ is the three-body energy in the
center of mass frame).

The integration in (\ref{apb1}) over $\bm{P}$ can be made trivially thanks to the momentum delta function, 
and the phase space can be rewritten as:
\begin{equation}
\Phi=\left( \frac{\mu_x}{m} \right)^{3/2} \left( \frac{\mu_y}{m} \right)^{3/2}
\int \kappa^5 d\kappa d\Omega_\kappa \delta\left(E-\frac{\hbar^2 \kappa^2}{2m}\right),
\label{apb2}
\end{equation}
where we have used that according to Eqs.(\ref{eq3}) and (\ref{eq4}) 
\begin{equation}
d\bm{p}_x d\bm{p}_y=\left( \frac{\mu_x}{m} \right)^{3/2} \left( \frac{\mu_y}{m} \right)^{3/2}
d\bm{k}_x d\bm{k}_y
\end{equation}
and that $d\bm{k}_x d\bm{k}_y = \kappa^5 d\kappa d\Omega_\kappa$, where $\Omega_\kappa$ represents
the hyperangles in momentum space.

The delta function in Eq.(\ref{apb2}) can be rewritten as:
\begin{equation}
\delta\left(E-\hbar^2 \frac{\kappa^2}{2m}\right)=
\frac{m}{\hbar^2 \kappa}\delta\left(
\kappa - \frac{\sqrt{2mE}}{\hbar}
\right),
\end{equation}
which means that, after integration over $\kappa$, the phase space (\ref{apb2}) takes the final form:
\begin{equation}
d^5\Phi=\left( \frac{\mu_x}{m} \right)^{3/2} \left( \frac{\mu_y}{m} \right)^{3/2}
 \kappa^4  \frac{m}{\hbar^2} d\Omega_\kappa .
\label{apb5}
\end{equation}

\subsection{Lab frame}

Let us consider now the 1+2 reaction in the lab frame, where the momentum of the target (the dimer) is zero,
and let us write the phase space $\Phi$ but using the momenta $\bm{p}_i$, $\bm{p}_j$, and 
$\bm{p}_k$ of the three particles after the breakup. Imposing energy and momentum conservation the 
phase space takes the form:
\begin{eqnarray}
\Phi&=&\int d\bm{p}_i d\bm{p}_j d\bm{p}_k 
\delta\left(E_{lab}
             -\frac{\hbar^2 p_i^2}{2m} -\frac{\hbar^2 p_j^2}{2m} -\frac{\hbar^2 p_k^2}{2m}\right)
 \nonumber \\  & &
\times \delta^{3}(\bm{p}_y^{(lab)}-\bm{p}_i-\bm{p}_j-\bm{p_k}),
\end{eqnarray}
where $\bm{p}_y^{(lab)}$ is the momentum of the incoming particle in lab frame, and $E_{lab}$ is the 
total three-body energy in the lab frame:
\begin{equation}
E_{lab}=E_{in}^{(lab)}+E_d=\frac{\hbar^2 p_y^{(lab)2}}{2m}+E_d,
\label{apb7}
\end{equation}  
where $E_{in}^{(lab)}$ is the energy of the projectile in the lab frame and 
$E_d$ is the binding energy of the dimer ($E_d<0$). 
For the particular case of three particles with equal mass, the incident momentum
in the lab frame is related to the one in the center of mass frame, $\bm{p}_y$,
by the simple relation $\bm{p}_y^{(lab)}=3\bm{p}_y/2$.

Making use of the momentum delta function we integrate now over $\bm{p}_k$, which leads to:
\begin{equation}
\Phi=\int d\bm{p}_i d\bm{p}_j  
\delta\left(f(\bm{p}_i,\bm{p}_j)\right)
\label{apb8}
\end{equation}
where
\begin{equation}
f(\bm{p}_i,\bm{p}_j)= E_{lab}
-\frac{\hbar^2 p_i^2}{2m} -\frac{\hbar^2 p_j^2}{2m} 
-\frac{\hbar^2(\bm{p}_y^{(lab)}-\bm{p}_i-\bm{p}_j )^2 }{2m}
\end{equation}

Using Eq.(\ref{apb7}) the function above can be written as:
\begin{eqnarray}
f(\bm{p}_i,\bm{p}_j)&=& E_d
-\frac{\hbar^2 p_i^2}{m} -\frac{\hbar^2 p_j^2}{m}
+
\label{apb10}\\ & &
+ \frac{\hbar^2}{m}\left(
p_y^{(lab)}p_i\mu_i+p_y^{(lab)}p_j\mu_j-p_ip_j \mu
\right)
\nonumber
\end{eqnarray}
with
\begin{eqnarray}
\mu_i&=&\frac{\bm{p}_y^{(lab)}\cdot\bm{p}_i}{p_y^{(lab)} p_i}=\cos{\theta_{p_i}} \label{apb11} \\
\mu_j&=&\frac{\bm{p}_y^{(lab)}\cdot\bm{p}_j}{p_y^{(lab)} p_j}=\cos{\theta_{p_j}} \label{apb12} \\
\mu&=&\frac{\bm{p}_i\cdot\bm{p}_j}{p_i p_j}=\sin{\theta_{p_i}}\sin{\theta_{p_j}} \cos \Delta\varphi+
                                            \mu_i \mu_j, \label{apb13}
\end{eqnarray}
where $\theta_{p_i}$ and $\theta_{p_j}$ are the polar angles associated to the direction of 
$\bm{p}_i$ and $\bm{p}_j$, respectively, and $\Delta \varphi$ is the difference between the
corresponding azimuthal angles. The $z$-axis is chosen along the momentum $\bm{p}_y^{(lab)}$.

We shall now exploit the well known property of the delta function:
\begin{equation}
\delta(g(x))=\sum_i \frac{1}{|g'(x_i)|}\delta(x-x_i)
\label{apb14}
\end{equation}
where the summation is over all the $x_i$ values satisfying $g(x_i)=0$, and where $g'(x_i)$ is the value
of the first derivative of $g(x)$ at the point $x_i$. Getting thus the derivative of Eq.(\ref{apb10}) with 
respect to $p_i$, and using Eq.(\ref{apb14}), it is then easy to rewrite $\delta(f(\bm{p}_i,\bm{p}_j))$ in
a more convenient form and integrate Eq.(\ref{apb8}) over $p_i$, which leads to:
\begin{equation}
\Phi=\frac{m}{\hbar^2} \int p_j^2 dp_j d\hat{\bm{p}}_i d\hat{\bm{p}}_j 
 \frac{p_i^2}{|2p_i-p_y^{(lab)}\mu_i+p_j\mu|},
\label{apb15}
\end{equation}
where for each value of $p_j$ and the directions $\hat{\bm{p}}_i$ and $\hat{\bm{p}}_j$, $p_i$ is obtained
as the solution of 
\begin{eqnarray}
 & & E_d
-\frac{\hbar^2 p_i^2}{m} -\frac{\hbar^2 p_j^2}{m}
+
\label{apb16b}\\ & &
+ \frac{\hbar^2}{m} \left(
p_y^{(lab)}p_i\mu_i+p_y^{(lab)}p_j\mu_j-p_ip_j \mu
\right)=0.
\nonumber
\end{eqnarray}

In the same way it is possible to integrate Eq.(\ref{apb8}) over $p_j$ instead of $p_i$, and get:
\begin{equation}
\Phi=\frac{m}{\hbar^2} \int p_i^2 dp_i d\hat{\bm{p}}_i d\hat{\bm{p}}_j 
 \frac{p_j^2}{|2p_j-p_y^{(lab)}\mu_j+p_i\mu|}.
\label{apb16}
\end{equation}

From Eqs.(\ref{apb15}) and (\ref{apb16}) it is then clear that:
\begin{equation}
\frac{dp_j}{|2p_i-p_y^{(lab)}\mu_i+p_j\mu|}=\frac{dp_i}{|2p_j-p_y^{(lab)}\mu_j+p_i\mu|}
\end{equation}

Since the energy $E_i$ of particle $i$ is given by $E_i=\hbar^2p_i^2/2m$ we have that
$dE_i=\hbar^2 p_i dp_i/m$ (and the same for particle $j$), and the expression above can be written as:
\begin{equation}
\frac{p_i dE_j}{|2p_i-p_y^{(lab)}\mu_i+p_j\mu|}=\frac{p_j dE_i}{|2p_j-p_y^{(lab)}\mu_j+p_i\mu|},
\end{equation}
or, in other words:
\begin{equation}
dE_j= \frac{p_j |2p_i-p_y^{(lab)}\mu_i+p_j\mu| }{p_i |2p_j-p_y^{(lab)}\mu_j+p_i\mu|} dE_i.
\label{apb21}
\end{equation}

We now define the arclength $S$ such that:
\begin{equation}
dS=\sqrt{(dE_i)^2+(dE_j)^2},
\label{apb21b}
\end{equation}
which from Eq.(\ref{apb21}) leads to:
\begin{eqnarray}
\lefteqn{
dE_i=} \label{apb22}
\\ & & \hspace*{-5mm}
\frac{p_i|2p_j-p_y^{(lab)}\mu_j+p_i\mu| dS}{
\left[
p_i^2 (2p_j-p_y^{(lab)}\mu_j+p_i\mu)^2 +p_j^2 (2p_i-p_y^{(lab)}\mu_i+p_j\mu)^2
\right]^{1/2}
} 
\nonumber
\end{eqnarray}

Finally, keeping in mind that $E_i=\hbar^2 p_i^2/2m$ and therefore $p_i dp_i=(m/\hbar^2) dE_i$, we can write
the phase space Eq.(\ref{apb16}) as:
\begin{equation}
\Phi=\left(\frac{m}{\hbar^2}\right)^2 \int p_i dE_i d\hat{\bm{p}}_i d\hat{\bm{p}}_j 
 \frac{p_j^2}{|2p_j-p_y^{(lab)}\mu_j+p_i\mu|},
\label{apb23}
\end{equation}
which by use of Eq.(\ref{apb22}) leads to the final expression for the phase space in the lab frame:
\begin{equation}
d^5\Phi=\left(\frac{m}{\hbar^2}\right)^2 K_S dS d\hat{\bm{p}}_i d\hat{\bm{p}}_j ,
\label{apb24}
\end{equation}
where
\begin{eqnarray}
\lefteqn{
K_S=}
\label{apb25} \\ & & \hspace*{-5mm}
\frac{p_i^2 p_j^2}{
\left[
p_i^2 (2p_j-p_y^{(lab)}\mu_j+p_i\mu)^2 +p_j^2 (2p_i-p_y^{(lab)}\mu_i+p_j\mu)^2
\right]^{1/2}
}
\nonumber
\end{eqnarray}

\section{Calculation of $p_i$ and $p_j$ for a given input}
\label{app3}

In this work the $n-d$ breakup reaction is investigated taking as input the incident energy
of the projectile in the lab frame $E_{in}^{(lab)}$, the polar angles $\theta_{p_i}$ and 
$\theta_{p_j}$ of the two neutrons observed after the breakup, and the difference between the
corresponding two azimuthal angles $\Delta \varphi$. 

From the energy $E_{in}^{(lab)}$ we can easily obtain the incident momentum of the projectile in the
lab frame ($p_y^{(lab)}$), the incident momentum in the center of mass frame ($p_y^{(in)}$) and the 
three-body momentum $\kappa$ (see Eq.(\ref{eq14}) and below). 

The angles $\theta_{p_i}$, $\theta_{p_j}$, and $\Delta \varphi$ permit us to obtain $\mu_i$, 
$\mu_j$, and $\mu$, which are given by Eqs.(\ref{apb11}), (\ref{apb12}), and (\ref{apb13}), 
respectively.

The remaining point is to determine for each value of the arclength $S$ the values of $p_i$ and $p_j$,
such that we can obtain $K_S$ according to Eq.(\ref{apb25}), and then obtain the cross section 
(\ref{eq13}) as a function of $S$. 

The momenta $p_i$ and $p_j$ are not independent. They are related through the expression (\ref{apb16b}),
which can be written as:
\begin{equation}
\alpha^2+p_i^2+p_j^2-p_y^{(lab)} p_i \mu_i - p_y^{(lab)} p_j \mu_j + p_i p_j \mu =0,
\end{equation}
where $\alpha^2=-mE_d/\hbar^2$ ($E_d<0$).

After some algebra the equation above can be seen to describe an ellipse, whose equation is given by:
\begin{equation}
\frac{(p'_i-p^C_i)^2}{a_i^2}+\frac{(p'_j-p_j^C)^2}{a_j^2}=1,
\label{apc2}
\end{equation}
where
\begin{eqnarray}
p^C_i&=& \frac{p_y^{(lab)}}{2\sqrt{2}}\frac{\mu_i-\mu_j}{1-\mu/2} \\
p^C_j&=& \frac{p_y^{(lab)}}{2\sqrt{2}}\frac{\mu_i+\mu_j}{1+\mu/2} ,
\end{eqnarray}
and $a_i$ and $a_j$ are given as
\begin{eqnarray}
a_i&=& \frac{\beta}{\sqrt{1-\frac{\mu}{2}}}, \\
a_j&=& \frac{\beta}{\sqrt{1+\frac{\mu}{2}}}, 
\end{eqnarray}
with
\begin{equation}
\beta^2=\frac{p_y^{(lab)2}}{8}\left[
\frac{(\mu_i-\mu_j)^2}{1-\frac{\mu}{2}}+\frac{(\mu_i+\mu_j)^2}{1+\frac{\mu}{2}}
\right] - \alpha^2
\end{equation}

Note that given the input, $p_i^C$, $p_j^C$, $a_i$, and $a_j$ are just numbers.

Finally, $p'_i$ and $p'_j$ are such that:
\begin{eqnarray}
p_i&=&\frac{p'_i+p'_j}{\sqrt{2}}, \label{apc8}\\
p_j&=&\frac{p'_j-p'_i}{\sqrt{2}}. \label{apc9} 
\end{eqnarray}

The equation of the ellipse (\ref{apc2}) can be written in parametric form as:
\begin{eqnarray}
p'_i-p_i^C &=&a_i \cos{\tilde{\varphi}}, \\
p'_j-p_j^C &=&a_j \sin{\tilde{\varphi}},
\end{eqnarray}
which permit to write Eqs.(\ref{apc8}) and (\ref{apc9}) as:
\begin{eqnarray}
p_i&=&\frac{a_i\cos{\tilde{\varphi}}+p_i^C+a_j\sin{\tilde{\varphi}}+p_j^C}{\sqrt{2}}, \label{apc12}\\
p_j&=&\frac{a_j\sin{\tilde{\varphi}}+p_j^C-a_i\cos{\tilde{\varphi}}-p_i^C}{\sqrt{2}}, \label{apc13} 
\end{eqnarray}
from which we get:
\begin{eqnarray}
dp_i&=&\frac{-a_i\sin{\tilde{\varphi}}+a_j\cos{\tilde{\varphi}}}{\sqrt{2}} d\tilde{\varphi}, \\
dp_j&=&\frac{a_j\cos{\tilde{\varphi}}+a_i\sin{\tilde{\varphi}}}{\sqrt{2}} d\tilde{\varphi}. 
\end{eqnarray}

Reminding now that $dE_i=\hbar^2 p_i dp_i/m$, and making use of the two equations above, we can
write the arclength $S$ given in Eq.(\ref{apb21b}) as:
\begin{equation}
dS=F(\tilde{\varphi})d\tilde{\varphi},
\end{equation}
where
\begin{eqnarray}
\lefteqn{
F(\tilde{\varphi})=}  \label{apc17}\\ & & \hspace*{-5mm}
\frac{\hbar^2}{m} \frac{1}{\sqrt{2}}
\left[
p_i^2(a_j\cos{\tilde{\varphi}}-a_i\sin{\tilde{\varphi}})^2+
p_j^2(a_j\cos{\tilde{\varphi}}+a_i\sin{\tilde{\varphi}})^2
\right]^{1/2}, \nonumber
\end{eqnarray}
where $p_i$ and $p_j$ are given as a function of $\tilde{\varphi}$ by Eqs.(\ref{apc12}) and 
(\ref{apc13}), respectively.

Therefore, the arclength $S$ can be obtained as a function of $\tilde{\varphi}$ from the expression:
\begin{equation}
S(\tilde{\varphi})=\int_{\tilde{\varphi}_0}^{\tilde{\varphi}} F(\tilde{\varphi}') d\tilde{\varphi}',
\label{apc18}
\end{equation}
where, for all the cases except case 3, $\tilde{\varphi}_0$ is defined such
that for $\tilde{\varphi}=\tilde{\varphi}_0$ then $p_j=0$. For case 3, since
the ellipse in Eq.(\ref{apc2}) does not cross the $p_j=0$ axis,
$\tilde{\varphi}_0$ is defined such that for $\tilde{\varphi}=\tilde{\varphi}_0$ then $p_i=0$.

In a practical case, given a value of $S$, one has to determine the angle $\tilde{\varphi}$ such that
$S(\tilde{\varphi})=S$. With this value of $\tilde{\varphi}$ the values of $p_i$ and $p_j$ are immediately 
obtained from Eqs.(\ref{apc12}) and (\ref{apc13}).

\section{Adiabatic expansion of three-body continuum wave functions}
\label{app4}

In the appendix of Ref.\cite{dan04} the authors give the general solution of the coupled-channel
problem for three-body continuum states in the presence of interaction potentials. 
Keeping the notation used in that reference, the continuum three-body wave function at a given 
three-body energy $E=\kappa^2 \hbar^2/2m$ is written as:
\begin{eqnarray}
\Psi&=&\sum_{JM} \frac{(2\pi)^3}{(\kappa \rho)^{5/2}} \sum_{K\gamma,K'\gamma'} 
\psi^J_{K\gamma,K'\gamma'}(\kappa \rho) \Upsilon_{JM}^{K\gamma}(\Omega_\rho) 
\nonumber \\ & &
\times \sum_{M'_L M'_S} \langle L' M'_L S' M'_S | J M \rangle
{\cal Y}_{K'L'M'_L}^{\ell'_x\ell'_y}(\Omega_\kappa)^*,
\label{eq3b}
\end{eqnarray}
where the functions ${\cal Y}_{K'L'M'_L}^{\ell'_x\ell'_y}$ are the usual hyperspherical harmonics, 
the index $\gamma$ collects the quantum numbers $\{\ell_x,\ell_y,L,S\}$, and $\Upsilon_{JM}^{K\gamma}$
is defined as the coupling between the hyperspherical harmonic ${\cal Y}_{KLM_L}^{\ell_x\ell_y}$ and
the three-body spin function $\chi_{S M_S}$:
\begin{equation}
\Upsilon_{JM}^{K\gamma}(\Omega_\rho)=
\sum_{M_L M_S} \langle L M_L S M_S | J M \rangle {\cal Y}_{KLM_L}^{\ell_x\ell_y}(\Omega_\rho) 
\chi_{SM_S}.
\label{defup}
\end{equation}

It is convenient to write the total three-body wave function as $\Psi=\sum_{S' M'_S} \Psi_{S' M'_S}$,
where:
\begin{eqnarray}
\Psi_{S'M'_S}&=&\sum_{JM} \frac{(2\pi)^3}{(\kappa \rho)^{5/2}} \sum_{K\gamma} \sum_{K'\ell'_x \ell'_y L'}
\psi^J_{K\gamma,K'\gamma'}(\kappa \rho) \Upsilon_{JM}^{K\gamma}(\Omega_\rho) 
\nonumber \\ & &
\times \sum_{M'_L} \langle L' M'_L S' M'_S | J M \rangle
{\cal Y}_{K'L'M'_L}^{\ell'_x\ell'_y}(\Omega_\kappa)^*,
\label{proj}
\end{eqnarray}
which represents the continuum three-body wave function where the spin function of the incoming 
channel is described by the quantum numbers $S'$ and $M'_S$. 

This can be easily seen because in the case of no interaction between the particles, the radial wave
functions $\psi^J_{K\gamma,K'\gamma'}(\kappa \rho)$ reduce to 
$i^K \sqrt{\kappa\rho}J_{K+2}(\kappa \rho)\delta_{KK'}\delta_{\gamma,\gamma'}$, and the three-body 
wave function becomes:
\begin{equation}
\Psi_{S'M'_S} \stackrel{\mbox{\tiny no interaction}}{\longrightarrow} e^{i(\bm{k}_x\cdot \bm{x}+\bm{k}_y\cdot \bm{y})} 
\chi_{S' M'_S},
\label{pw}
\end{equation}
which is a three-body plane wave multiplied by the three-body spin function in the incoming channel.

Using now the expression Eq.(\ref{defup}) the wave function in Eq.(\ref{proj}) can be written in a more 
compact way as:
\begin{eqnarray}
\Psi_{S'M'_S} &=&  \sum_{JM} \frac{(2\pi)^3}{(\kappa \rho)^{5/2}}     
\sum_{K\gamma} \sum_{K' \ell'_x \ell'_y L'}
\psi^J_{K\gamma,K'\gamma'}(\kappa \rho) \Upsilon_{JM}^{K\gamma}(\Omega_\rho) 
\nonumber \\ & &
\times \langle \chi_{S'M'_S}|\Upsilon_{JM}^{K'\gamma'}(\Omega_\kappa)\rangle^*.
\label{eq3b2}
\end{eqnarray}

The transformation of the wave function $\Psi_{S'M'_S}$ from the HH basis to the basis formed by 
the hyperangular functions $\Phi^{JM}_n$ defined in Eq.(\ref{eq16}) can be easily made 
through the relation
\begin{equation}
\Upsilon_{JM}^{K \gamma}=\sum_{n} \langle \Phi^{JM}_n | 
                     \Upsilon_{JM}^{K \gamma} \rangle \Phi^{JM}_n,
\end{equation}
which leads to:
\begin{eqnarray}
\Psi_{S'M'_S} & = & \sum_{JM} \frac{(2\pi)^3}{(\kappa \rho)^{5/2}}     
\\ & & \hspace*{-2cm}\times
\sum_{n n'} f_{n n'}^{S' J}(\kappa \rho) \Phi^{JM}_n(\rho,\Omega_\rho)
\langle \chi_{S'M'_S}|\Phi^{JM}_{n'}(\kappa, \Omega_\kappa)\rangle^*,
\nonumber
\end{eqnarray}
and where we have defined:
\begin{eqnarray}
\lefteqn{f_{nn'}^{S'J}(\kappa \rho)=}
\nonumber \\ & &
\sum_{K\gamma} \sum_{K' \ell'_x \ell'_y L'}
\psi^J_{K\gamma,K'\gamma'}(\kappa \rho) 
\langle \Phi^{JM}_n(\rho, \Omega_\rho) |\Upsilon_{JM}^{K\gamma}(\Omega_\rho)\rangle
\nonumber \\ & &
\langle\Upsilon_{JM}^{K'\gamma'}(\Omega_\kappa)|\Phi^{JM}_{n'}(\kappa, \Omega_\kappa) \rangle.
\end{eqnarray}

As shown in Eq.(\ref{pw}), the expansion in Eq.(\ref{eq3b}) has been written in such a way that in case of
no interaction between the particles the wave function Eq.(\ref{proj}) reduces basically to  
a three-body plane.  In fact, the way they are written, Eqs.(\ref{eq3b}) or (\ref{proj}) are appropriate 
to describe $3\rightarrow 3$ reactions, and would permit to extract the corresponding $3\rightarrow 3$ 
transition amplitude.

However, when dealing with 1+2 reactions, in case of no interaction between the projectile and the dimer,
the three-body continuum wave function must reduce to 
\begin{equation}
\Psi_{\sigma_d \sigma_p} \stackrel{\mbox{\tiny no interaction}}{\longrightarrow} 
e^{i\bm{k}_y\cdot \bm{y}} \chi_{s_d \sigma_d} \chi_{s_p \sigma_p},
\label{asym2}
\end{equation}
where $\chi_{s_d \sigma_d}$ is the dimer
wave function (with spin $s_d$ and projection $\sigma_d$), and $\chi_{s_p \sigma_p}$ is the spin 
function of the projectile. As we can see, the three-body plane wave is now an ordinary two-body
plane wave (describing the relative free motion between the projectile and the dimer center of mass).
In order to obtain such two-body plane wave with the correct normalization the factor $(2 \pi)^3$  
in the expansion in Eq.(\ref{eq3b}), and therefore also in the following expressions of the
continuum wave function, has to be replaced by $(2\pi)^{3/2}$.

Also, as seen from Eq.(\ref{asym2}), if we want to describe the incoming spin state by 
$\chi_{s_d \sigma_d} \chi_{s_p \sigma_p}$ (instead of the total three-body spin and projection $S'$ and
$M'_S$), 
the corresponding wave function $\Psi_{\sigma_d\sigma_p}$ can be obtained from $\Psi_{S'M'_S}$ through
the simple relation:
\begin{equation}
\Psi_{\sigma_d \sigma_p}=\sum_{S' M'_S} \langle \chi_{S'M'_S} | \sigma_p \sigma_d \rangle \Psi_{S' M'_S},
\end{equation}
where $|\sigma_p \sigma_d \rangle$ represents the spin state 
$|\chi_{s_d \sigma_d} \chi_{s_p \sigma_p}\rangle$, and from which we can write:
\begin{eqnarray}
\lefteqn{ \Psi_{\sigma_d \sigma_p}  =  }
\label{finfun}	\\ & & \hspace*{-3mm}
\sum_{JM} \frac{(2\pi)^{3/2}}{(\kappa \rho)^{5/2}}     
\sum_{n n'} f_{n n'}^{J}(\kappa \rho) \Phi^{JM}_n(\rho, \Omega_\rho)
\langle \sigma_d \sigma_p |\Phi^{JM}_{n'}(\kappa, \Omega_\kappa)\rangle^*,
\nonumber
\end{eqnarray}
where
\begin{equation}
f_{n n'}^{J}(\kappa \rho)=\sum_{S' M'_S} 
\langle \chi_{S'M'_S} | \sigma_d \sigma_p\rangle f_{nn'}^{S'J}(\kappa \rho) 
\langle \sigma_d \sigma_p| \chi_{S'M'_S} \rangle, 
\end{equation}
and where we have assumed that the final radial wave functions do not depend on the spin projections, and 
that the interaction does not mix different spin states $|\sigma_d \sigma_p \rangle$.

As usual, in Eq.(\ref{finfun}) the indices $n$ and $n^\prime$ refer to the outgoing and incoming channels,
respectively, and Eq.(\ref{finfun}) gives the full three-body wave function, where all the possible 
incoming and outgoing channels are contained. Only the spin projection in the incoming channel are
specified. If we now consider a specific incoming channel, 
let us say channel 1, the summation over $n'$ disappears, and $n'=1$. Also, if we want to restrict ourselves
to outgoing breakup channels, the summation over $n$ should run only over the adiabatic terms
associated to breakup of the dimer. In particular, for the case in which only one 1+2 channel
exist (the incoming channel 1), we have that the summation over $n$ runs over all $n>1$.

Finally, if we project over a specific spin state $|\sigma_i \sigma_j \sigma_k \rangle$ for the 
three particles after the breakup, we then get the final expression for the adiabatic expansion 
of the three-body wave function describing the breakup of the dimer, and it is given by:
\begin{eqnarray}
\Psi_{\sigma_d \sigma_p}^{\sigma_i \sigma_j \sigma_k} & = & 
              \sum_{JM} \frac{(2\pi)^{3/2}}{(\kappa \rho)^{5/2}} 
\sum_{n>1} f_{n 1}^{J}(\kappa \rho) \label{expan}
\\ & &
\langle \sigma_i \sigma_j \sigma_k |\Phi^{JM}_n(\rho, \Omega_\rho) \rangle
\langle \sigma_d \sigma_p |\Phi^{JM}_{n'}(\kappa, \Omega_\kappa)\rangle^*.
\nonumber
\end{eqnarray}

\section{Asymptotic incoming angular eigenfunction in momentum space}
\label{app5}

In coordinate space, the angular eigenfunction $\Phi_1(\rho,\Omega_\rho)$ associated to a
1+2 channel behaves for large values of $\rho$ as \cite{nie01}:
\begin{eqnarray}
\lefteqn{
\Phi_1^{JM}(\rho,\Omega_\rho)\stackrel{\rho\rightarrow \infty}{\rightarrow}
 } \label{apd1} \\ & &
        \left(\frac{m}{\mu_x} \right)^{3/4}
        \rho^{3/2} \left[ \psi_d^{j_x}(\bm{r}_x) 
        \otimes \left[Y_{\ell_y}(\Omega_y) \otimes \chi_{s_y} \right]^{j_y}\right]^{JM},
\nonumber
\end{eqnarray}
where $\psi_d^{j_x}(\bm{r}_x)$ is the dimer wave function, which is normalized to 1 in the 
relative coordinate $\bm{r}_x$ between the two particles in the dimer, and  
whose angular momentum is denoted by $j_x$. The projectile-dimer relative orbital angular
momentum, $\ell_y$, and the spin of the projectile, $s_y$, couple to the total angular momentum 
$j_y$, which in turn couples to $j_x$ to provide the total angular momentum $J$ of the three-body
system, whose projection is given by $M$. Finally, $\mu_x$ is the reduced mass of the two particles in
the dimer.

After a Fourier transformation the dimer wave function becomes $\psi_d^{j_x}(\bm{p}_x)$, 
also normalized to 1 in $\bm{p}_x$, where $\bm{p}_x$
is the relative momentum between the two particles in the dimer (Eq.(\ref{eq3})). Also, the polar
and azimuthal angles $\Omega_y$ describing the direction of the Jacobi coordinate $\bm{y}$ become
$\Omega_{k_y}$, which describe the direction of the incident Jacobi momentum $\bm{k}_y^{(in)}$ defined as in 
Eq.(\ref{eq4}). Therefore, the Fourier transform of Eq.(\ref{apd1}) must have the form:
\begin{equation}
\Phi_1^{JM}(\kappa,\Omega_\kappa)\rightarrow
        C \left[ \psi_d^{j_x}(\bm{p}_x) 
        \otimes \left[Y_{\ell_y}(\Omega_{k_y}) \otimes \chi_{s_y} \right]^{j_y}\right]^{JM},
  \label{apd2} 
\end{equation}
where $C$ is a normalization constant.

The value of $C$ can be obtained by imposing that:
\begin{equation}
\int d\Omega_\kappa \left| \Phi_1(\kappa,\Omega_\kappa)\right|^2 = 1,
\end{equation}
which after using the expansion
\begin{equation}
\Psi_d^{j_x m_x}(\bm{p}_x)=\sum_{\ell_x s_x} \Psi^{j_x}_{\ell_x s_x}(p_x)
\left[Y_{\ell_x}(\Omega_{k_x}) \otimes \chi_{s_x} \right]^{j_x m_x}
\end{equation}
leads to:
\begin{equation}
\left| C \right|^2
\int d\alpha_\kappa \sin^2 \alpha_\kappa \cos^2 \alpha_\kappa 
\sum_{\ell_x s_x} \left| \Psi^{j_x}_{\ell_x s_x}(p_x) \right|^2=1.
\end{equation}

Using now that $k_x=\kappa \sin \alpha_\kappa$ 
(and therefore $dk_x=\kappa \cos \alpha_\kappa d\alpha_\kappa$) and $k_y^{(in)}=\kappa \cos \alpha_\kappa$, 
the expression above can be rewritten as:
\begin{equation}
\left| C \right|^2 \frac{k_y^{(in)}}{\kappa^4}
\int dk_x k_x^2   \sum_{\ell_x s_x} \left| \Psi^{j_x}_{\ell_x s_x}(p_x) \right|^2=1,
\end{equation}
which, after taking into account that $ k_x^2 dk_x=(m/\mu_x)^{3/2} p_x^2 dp_x$, and keeping in mind that
$\psi_d^{j_x}(\bm{p}_x)$ is normalized to 1 in $\bm{p}_x$, leads to:
\begin{equation}
\left| C \right|^2=\frac{\kappa^4}{k_y^{(in)}} \left( \frac{\mu_x}{m} \right)^{3/2}, 
\end{equation}
and therefore:
\begin{eqnarray}
\lefteqn{
\Phi_1^{JM}(\kappa,\Omega_\kappa)\rightarrow
} \\ & & 
        \frac{\kappa^2}{\sqrt{k_y^{(in)}}} \left( \frac{\mu_x}{m} \right)^{3/4} 
        \left[ \psi_d^{j_x}(\bm{p}_x) 
        \otimes \left[Y_{\ell_y}(\Omega_{k_y}) \otimes \chi_{s_y} \right]^{j_y}\right]^{JM}.
\nonumber
\end{eqnarray}

In the particular case of only relative $s$-waves between the particles, the expression above reduces to:
\begin{eqnarray}
\lefteqn{
\Phi_1^{JM}(\kappa,\Omega_\kappa)\rightarrow
} \label{apd9} \\ & & 
        \frac{1}{\sqrt{4\pi}} \left( \frac{\mu_x}{m} \right)^{3/4} \frac{\kappa^2}{\sqrt{k_y^{(in)}}}  
        \left[ \psi_d^{s_d}(\bm{p}_x) 
        \otimes \chi_{s_p} \right]^{JM},
\nonumber
\end{eqnarray}
where we have recovered the notation of Section~\ref{amplitude}, where $s_d$ represents the spin of the dimer
and $s_p$ is the spin of the projectile. 

From Eq.(\ref{apd9}) it is now simple to see that the projection over the intrinsic states of projectile and
target, the asymptotic 
incoming angular wave function in momentum space is given by:
\begin{equation}
\langle \sigma_d \sigma_p | \Phi_1^*(\kappa,\Omega_\kappa)\rangle=
\frac{1}{\sqrt{4\pi}} \left( \frac{\mu_x}{m} \right)^{3/4} \frac{\kappa^2}{\sqrt{k_y^{(in)}}}
\langle s_d \sigma_d s_p \sigma_p | J M \rangle,
\label{apd10}
\end{equation}
which is the expression given in Eq.(\ref{eq22}).

\section{ Fresnel integrals}
\label{app6}
The aim of this section is to compute analytically the integral:
\begin{equation}
\int_{y_{\mbox{\scriptsize max}}}^\infty dy
\left( \frac{C_m \sin(\kappa y)\sin(k_y y)}{y^{m+3/2}} + 
       \frac{D_m \cos(\kappa y) \sin(k_y y)}{y^{m+3/2}} \right),
\label{eqf1}
\end{equation}
which can also be written as:
\begin{eqnarray}
\frac{C_m}{2} \int_{y_{\mbox{\scriptsize max}}}^\infty dy
 \frac{\cos((\kappa-k_y)y)-\cos((\kappa+k_y)y)}{y^{m+3/2}} -
\label{eqf2} \\  & &  \hspace*{-6cm}
\frac{D_m}{2} \int_{y_{\mbox{\scriptsize max}}}^\infty dy
 \frac{\sin((\kappa-k_y)y)-\sin((\kappa+k_y)y)}{y^{m+3/2}}.
\nonumber
\end{eqnarray}

Therefore, the calculation of the integrals above requires just the calculation of the integrals
of the type:
\begin{equation}
\int_{y_{\mbox{\scriptsize max}}}^\infty dy \frac{\cos(ay)}{y^{m+3/2}} \mbox{  and  } 
\int_{y_{\mbox{\scriptsize max}}}^\infty dy \frac{\sin(ay)}{y^{m+3/2}},
\label{eqf3}
\end{equation}
where $a$ can be either $(\kappa-k_y)$ or $(\kappa+k_y)$.

Taking now $t=ay$ the two integrals above take the form:
\begin{equation}
a^{m+1/2}\int_{t_{\mbox{\scriptsize max}}}^\infty dt \frac{\cos t}{t^{m+3/2}} \mbox{  and  } 
a^{m+1/2}\int_{t_{\mbox{\scriptsize max}}}^\infty dt \frac{\sin t}{t^{m+3/2}},
\label{eqf4}
\end{equation}
where $t_{\mbox{\scriptsize max}}=ay_{\mbox{\scriptsize max}}$.

Let us now define:
\begin{eqnarray}
\tilde{S}_\nu(z)&=&\int_z^\infty dx \frac{\sin x}{x^\nu} \label{eqf5}\\
\tilde{C}_\nu(z)&=&\int_z^\infty dx \frac{\cos x}{x^\nu},\label{eqf6} 
\end{eqnarray}
such that the two integrals in Eq.(\ref{eqf4}) are $\tilde{C}_{3/2}(t_{\mbox{\scriptsize max}})$
and $\tilde{S}_{3/2}(t_{\mbox{\scriptsize max}})$, respectively.

If we now take into account that:
\begin{eqnarray}
\frac{\sin x}{x^\nu}dx & = &  \frac{1}{\nu-1} \frac{\cos x}{x^{\nu-1}} dx 
    -\frac{1}{\nu-1} d\left( \frac{\sin x}{x^{\nu-1}}\right)
\\
\frac{\cos x}{x^\nu}dx & = & -\frac{1}{\nu-1} \frac{\sin x}{x^{\nu-1}} dx 
    -\frac{1}{\nu-1} d\left( \frac{\cos x}{x^{\nu-1}}\right),
\end{eqnarray}
we can the immediately get the recurrence relations:
\begin{eqnarray}
\tilde{S}_\nu(z) & = &  
    \frac{1}{\nu-1} \frac{\sin z}{z^{\nu-1}} + \frac{1}{\nu-1} \tilde{C}_{\nu-1}(z)
\label{eqf9}
\\
\tilde{C}_\nu(z) & = &  
    \frac{1}{\nu-1} \frac{\cos z}{z^{\nu-1}} - \frac{1}{\nu-1} \tilde{S}_{\nu-1}(z),
\label{eqf10}
\end{eqnarray}

Therefore, knowledge of $\tilde{S}_{1/2}(z)$ and $\tilde{C}_{1/2}(z)$ would permit, through the
recurrence relations in Eqs.(\ref{eqf9}) and (\ref{eqf10}) to obtain the integrals in Eq.(\ref{eqf3}),
and therefore the integral in Eq.(\ref{eqf2}) (or (\ref{eqf1})).

The integrals $\tilde{S}_{1/2}(z)$ and $\tilde{C}_{1/2}(z)$ are known analytically, and they are 
given by:
\begin{eqnarray}
\tilde{S}_{1/2}(z)&=&\sqrt{\frac{\pi}{2}}-\frac{\sqrt{\pi}}{2}
\left(
\sqrt{i} \mbox{ erf}(\sqrt{iz}) + \sqrt{-i} \mbox{ erf}(\sqrt{-iz})
\right)  \nonumber \\
\tilde{C}_{1/2}(z)&=&\sqrt{\frac{\pi}{2}}-\frac{\sqrt{\pi}}{2}
\left(
\sqrt{-i} \mbox{ erf}(\sqrt{-iz}) + \sqrt{i} \mbox{ erf}(\sqrt{iz})
\right)  \nonumber ,
\end{eqnarray}
where ``erf'' stands for the usual error function.

\end{document}